\documentclass{svmult}

\usepackage{graphicx}			% package to include pdf graphics
\usepackage{fancyhdr}			% gives headers see below
\usepackage{geometry}			% package to give easy control of paper and margin size
\usepackage{amsmath}			% packages to give lots of maths stuff
\usepackage{amssymb}
\usepackage{latexsym}
\usepackage{subcaption} %Ídem que Caption pero con subfiguras
\usepackage{float} %Permite manejar/mover/partir mejor figuras y tablas
\usepackage{braket}
\usepackage{bbold} % allows us to use the command \mathbb
\usepackage{mathtools}
\usepackage{physics}
\usepackage{dsfont}
\usepackage{tikz-feynman}
\usepackage{listings}
\usepackage{bm}
\usepackage{dirtytalk}

\usepackage{mcite}
\usepackage{cite}

\usepackage[hyperfootnotes=false,colorlinks=false]{hyperref}
\hypersetup{%backref,
    colorlinks,%
    citecolor=black,%
    filecolor=black,%
    linkcolor=black,%
    urlcolor=cyan,
    }
\usepackage{url}
\usepackage{color}
\usepackage{setspace}
\usepackage{inputenc}
\usepackage[english]{babel}

\usepackage{xcolor}

\renewcommand{\[}{\begin{equation}}
	\renewcommand{\]}{\end{equation}}
\def\beq{\begin{equation}}
	\def\eeq{\end{equation}}
\newcommand{\be}{\begin{eqnarray}}
	\newcommand{\ee}{\end{eqnarray}}

\renewcommand{\texttt}{{}}

\def\bs{\begin{subequations}}
	\def\es{\end{subequations}}

\def\Bc{\mathcal{B}}
\def\Cc{\mathcal{C}}
\def\Dc{\mathcal{D}}

\def\Fc{\mathcal{F}}

\def\Lc{\mathcal{L}}
\def\Mc{\mathcal{M}}
\def\Nc{\mathcal{N}}
\def\Oc{\mathcal{O}}
\def\Pc{\mathcal{P}}

\def\Rc{\mathcal{R}}

\def\Tc{\mathcal{T}}

\def\Wc{\mathcal{W}}

\newcommand{\tia}[1]{}

\newcommand{\bea}{\begin{eqnarray}}
	\newcommand{\eea}{\end{eqnarray}}
\newcommand{\beas}{\begin{eqnarray*}}
	\newcommand{\eeas}{\end{eqnarray*}}
\newcommand{\bal}{\begin{aligned}}
	\newcommand{\eal}{\end{aligned}}

\def\({\left(}
\def\){\right)}

\newcommand{\LF}{\left(}
\newcommand{\RF}{\right)}
\newcommand{\LT}{\left[}
\newcommand{\RT}{\right]}

\newcommand{\pd}{\partial}

\DeclareFontFamily{U}{mathb}{\hyphenchar\font45}
\DeclareFontShape{U}{mathb}{m}{n}{
      <5> <6> <7> <8> <9> <10> gen * mathb
      <10.95> mathb10 <12> <14.4> <17.28> <20.74> <24.88> mathb12
      }{}
\DeclareSymbolFont{mathb}{U}{mathb}{m}{n}
\DeclareFontSubstitution{U}{mathb}{m}{n}

\let\dot\relax
\DeclareMathAccent{\dot}{0}{mathb}{"39}
\let\ddot\relax
\DeclareMathAccent{\ddot}{0}{mathb}{"3A}
\let\dddot\relax
\DeclareMathAccent{\dddot}{0}{mathb}{"3B}
\let\ddddot\relax
\DeclareMathAccent{\ddddot}{0}{mathb}{"3C}

\title*{Cosmology in nonlocal gravity}

\author{Alexey S. Koshelev\thanks{corresponding author}, K. Sravan Kumar and Alexei A. Starobinsky{}}
\institute{Alexey S. Koshelev \at School of Physiical Science and Thechnology, ShanghaiTech University, 201210 Shanghai, China\\Departamento de F\'isica, Centro de Matem\'atica e Aplica\c{c}oes (CMA-UBI),
  Universidade da Beira Interior, 6200 Covilh\~a, Portugal \\ \email{askoshelev@shanghaitech.edu.cn} \and
K. Sravan Kumar \at Institute of Cosmology \& Gravitation,
	University of Portsmouth,
	Dennis Sciama Building, Burnaby Road,
	Portsmouth, PO1 3FX, United Kingdom\\ \email{sravan.kumar@port.ac.uk} \and Alexei A. Starobinsky \at L. D. Landau Institute for Theoretical Physics RAS, Chernogolovka, Moscow region 142432,
Russian Federation\\  Kazan Federal University, Kazan 420008, Republic of Tatarstan, Russian
Federation\\ \email{alstar@landau.ac.ru}}

\begin{document}

\maketitle

\abstract{
  In this chapter we review the recent developments of realizing $R^2$-like inflation in the framework of a most general UV nonlocal extension of Einstein's general theory of relativity (GR). It is a well-motivated robust approach towards quantum gravity. In the past decades, nonlocal gravitational theories which are quadratic in curvature have been understood to be ghost-free and super-renormalizable around maximally symmetric spacetimes. However, in the context of early Universe cosmology we show that one must go beyond the quadratic curvature nonlocal gravity in order to achieve a consistent ghost-free framework of Universe evolution from quasi de Sitter to Minkowski spacetime. In this regard, we discuss a construction of a most general nonlocal gravity action that leads to $R^2$-like inflation and discuss the corresponding observational predictions for the scalar and tensor spectral tilts, tensor-to-scalar ratio, and the primordial non-Gaussianities. We present an analysis of how the nonlocal inflationary cosmology goes beyond the established notions of effective field theories of inflation. Finally, we comment on some open questions and prospects of higher curvature nonlocal gravity on its way of achieving the UV completion. 
}

\newpage

 \section{Introduction}
 
Most often a prominent goal in the quantum gravity research is to achieve a short distance and time scale modification of Einstein's general theory of relativity (GR) and to find its ultraviolet (UV) completion via demanding Unitarity and renormalizability around Minkowski spacetime. However, when it comes to probing any theory of quantum gravity, one needs to look for the applications in the context of cosmology and black hole physics which implies the need of understanding quantum nature of gravity in curved and dynamical spacetime. This means UV completion of gravity should not only be restricted to Minkowski spacetime but also to the curved spacetime. The goal of obtaining a Unitary and renormalizable gravity theory around Minkowski background provides the essential steps we should take to modify GR, while the goal of explaining the early Universe and black hole physics produces practical restrictions on achieving a UV completion and also constrains the parameter space of the UV complete theory. 
The so far best example of a renormalizable gravity is represented by the Stelle's 4th order action \cite{Stelle:1977ry}
\begin{equation}
	S_{\rm 4} = \int d^4x\sqrt{-g} \Bigg[\frac{M_p^2}{2}R+\frac{f_0}{2} R^2+\frac{f_{0C}}{2}W_{\mu\nu\rho\sigma}W^{\mu\nu\rho\sigma}\Bigg]
	\label{Stelleact}
\end{equation}
where $f_0,\,f_{0C} $ are dimensionless coefficients and $M_p=\frac{1}{\sqrt{8\pi G}}$ is the reduced Planck mass. It is a renormalizable theory of gravity which is an extension of GR with additional Ricci scalar square ($R^2$) and Weyl tensor square ($W^2= W_{\mu\nu\rho\sigma}W^{\mu\nu\rho\sigma}$) in the action. However it is well-known that Stelle's gravity is non-Unitary due to the presence of an additional tensor ghost. On the other hand,  the model of Starobinsky inflation based on a quadratic scalar curvature extension of GR \cite{Starobinsky:1980te}  
\begin{equation}
	S^{\rm local}_{\rm R+R^2} = \int d^4x\sqrt{-g} \Bigg[\frac{M_p^2}{2}R+\frac{f_0 }{2}R^2\Bigg] 
	\label{Staroact}
\end{equation}
where $f_0=\frac{M_p^2}{6M^2}$ with $M$ being a mass of an additional scalar degree of freedom, 
is strongly supported by the latest observations of cosmic microwave background (CMB) by Planck satellite \cite{Akrami:2018odb} with the value of $M\approx 1.3\times 10^{-5} M_p$ for the $N=(50-60)$ number of e-foldings before the end of inflation. The success of Starobinsky inflation makes a strong point to the higher curvature modification of gravity, and in the context of Stelle gravity we must impose the inequality 
$f_0\gg f_{0C}$ in order to push the ghost away from affecting the primordial physics. Furthermore, the presence of $R^2$ term as an extension of GR is inevitable if one considers the contribution of  1-loop matter corrections to the graviton self energy \cite{Duff:1993wm}. Although, we do not know what is a theory of gravity at the Planck scale we can greatly expect the necessity of higher curvature terms \cite{Barvinsky:1994cg,Barvinsky:1990up,Barvinsky:1994hw}. In this sense the Weyl tensor square term is important for quantum gravity \cite{Shaposhnikov:2022zhj}, although we require $f_0\gg f_{0C}$ for the stability of inflation.\footnote{It is interesting to note here that the Weyl ghost can be avoided by using new quantum prescriptions \cite{Anselmi:2020lpp} or by considering the dressed propagator of graviton with a notion of unstable ghosts \cite{Donoghue:2021cza}. However,  further investigation is needed in the scope of quantum field theory in curved spacetime in order to check if the ghost problem can be fully alleviated.} To achieve a consistent theory of quantum gravity it is essential to go beyond the Stelle gravity but any finite derivative extension of it would obviously lead to the well-known Ostrogradsky instability \cite{Woodard:2015zca}. 

Several approaches towards quantum gravity do have a common feature known as nonlocality which arises due to an effect of infinite derivative terms in the action \cite{Koshelev:2020xby}. Considering nonlocality as a principle, a straightforward extension of Stelle gravity with an infinite number of covariant derivatives has been shown to resolve the ghost problem of the theory \cite{Krasnikov:1987yj,Tomboulis:1997gg,Biswas:2010zk,Modesto:2011kw} and has been projected to be a prominent candidate for a quantum gravity. The nonlocal quadratic curvature gravity (NLQG) theories in particular have been extensively studied in the past decade in the contexts of Unitarity, renormalizability with some applications in cosmology and astrophysics \cite{Buoninfante:2022ild,Biswas:2005qr,Koshelev:2020foq,Kolar:2021qox,Kolar:2021rfl,Koshelev:2013lfm,Koshelev:2018hpt,Koshelev:2017bxd}. The NLQG  is shown to be ghost-free around Minkowski spacetime (raising chances for it to be Unitary) and renormalizability of the theory is so far shown by the methods of power counting within the restricted class of  form factors which are analytic infinite derivative (AID) operators \cite{Tomboulis:1997gg,Modesto:2011kw,Tomboulis:2015esa,Tomboulis:2015gfa} (see also the recent review article \cite{BasiBeneito:2022wux} and references therein).\footnote{Also in some restricted class of nonlocal gravity actions renormalizability in the de Sitter and Anti-de Sitter backgrounds also has been achieved \cite{Koshelev:2017ebj}.} In the recent years, NLQG has been extensively studied in the scope of embedding $R+R^2$ or Starobinsky inflation in a UV complete framework \cite{Craps:2014wga,Koshelev:2016xqb,Koshelev:2017tvv,Koshelev:2020foq,Koshelev:2020xby}. Proven successful this embedding of $R^2$ inflation in NLQG has led to a nonlocal $R^2$-like inflation which has interesting observational consequences in the form of new predictions for the inflationary observables such as tensor-to-scalar ratio ($r$), tensor tilt ($n_t$) and primordial non-Gaussianities (PNGs) \cite{Koshelev:2020xby}. Therefore, future CMB and primordial gravitational wave (PGW) observations certainly will be able to shed light on signatures of inflation in nonlocal gravity \cite{Koshelev:2020xby}. As we discussed earlier, cosmological or astrophysical implications of any framework of quantum gravity will further anchor our understanding of quantum gravity. Indeed even though NLQG framework is successfully understood to be a super-renormalizable theory of quantum gravity, the cosmological application of it does demands the further extension of NLQG if we want to achieve a ghost-freeness simultaneously in inflationary and Minkowski backgrounds. This has been shown in the recent developments of generalized nonlocal $R^2$-like inflation \cite{Koshelev:2022olc,Koshelev:2022bvg}. It has been well-known that the ghost-free requirement of NLQG requires nonlocal form factors to be analytic functions not only of the d'Alembertian operator but also of background curvature \cite{Biswas:2016egy,Koshelev:2017ebj,SravanKumar:2019eqt,SravanKumar:2018dlo}. This implies that the NLQG is an incomplete framework and it could probably be a part of a more generic model. From the cosmological application point of view, NLQG has been extended to generalized nonlocal quantum gravity (GNLQG)  with higher curvature nonlocal terms, so that one can achieve a consistent regime of $R^2$-like inflation \cite{Koshelev:2022olc,Koshelev:2022bvg}. These developments are significant advances compared to the earlier known NLQG. This in turn  extends and sets new goals for our understanding of quantum gravity. In \cite{Koshelev:2022olc,Koshelev:2022bvg} a thorough study of $R^2$-like inflation in GNLQG has been performed and inflationary observables are derived bringing new insight into the physics of primordial Universe.\footnote{We note that in the context of a version of NLQG without any propagating scalar degree of freedom, alternative framework has been proposed recently which predicts $r\gtrsim 0.01$ \cite{Calcagni:2022tuz,Modesto:2022asj}. In this review, we stick ourselves to the discussion of early Universe inflationary cosmology rather to any of its adhoc alternatives. }

Inflationary cosmology so far has been intensively dominated by the numerous frameworks of scalar fields that are motivated from various approaches to UV complete physics such as string theory and supergravity \cite{Baumann:2018muz,Linde:2014nna,Martin:2013tda}. However, already in the pioneer paper \cite{Starobinsky:1980te} it had been demonstrated that a viable and full inflationary model including creation and heating of matter after inflation can be realized purely geometrically without introduction of scalar fields. Especially after the release of first Planck data \cite{Ade:2015lrj,Akrami:2018odb} there has been an increasing activity of finding extensions to the $R^2$-like framework combined with UV complete setups \cite{Kehagias:2013mya} and in this regard a possible UV completion of $R^2$ gravity within the scope of nonlocal gravity not only adds a new perspective to the current wide activities of inflationary cosmology but also highlights the possible role of nonlocality in the early Universe physics. Namely, the recent studies \cite{Koshelev:2022olc,Koshelev:2022bvg} specifically put these efforts on par to the contemporary attempts of realizing of the physics of inflation through the so called effective field theory (EFT) of a single field inflation or a multifield inflation \cite{Cheung:2007st,Senatore:2010wk}. To be more precise, the EFT of single field inflation (EFT-SI) prescribes that any deviation from a single field inflationary slow-roll consistency relation ($r=-8n_t$) requires the sound speeds of scalar and tensor degrees of freedom to deviate from Unity. However, the nonlocal $R^2$-like inflation both in NLQG and GNLQG \cite{Koshelev:2020foq,Koshelev:2022olc} violates the consistency relation $r=-8n_t$ preserving the Unit sound speeds of perturbed modes. Moreover, in the nonlocal $R^2$-like inflation we obtain both positive and negative values of tensor tilt which is entirely a new effect in the context of geometric formulations of inflation. Especially the possibilities of positive values for $n_t$ posits a counterexample to the alternative frameworks of inflation such as string gas cosmology \cite{Brandenberger:2015kga}. Furthermore, the studies of primordial non-gaussianities (PNG) in the nonlocal gravity inflation has revealed that one can obtain nontrivial shapes of PNGs that are otherwise thought to be not possible in standard EFT formulations of a single field and multifield inflation \cite{Koshelev:2020foq,Koshelev:2022bvg}. This is purely due to the nonlocal nature of gravity and it opens a new window for possible observational outcomes of inflation that should be kept in mind when interpreting results of 
%it does tells us a new picture of understanding inflation and it also does impacts a great caution in reading the physics of inflation from the 
future CMB and PGW probes such as such as CMBS4 \cite{CMB-S4:2020lpa}, LiteBIRD \cite{LiteBIRD:2022cnt}, CORE \cite{CORE:2016ymi} and e-LISA \cite{Ricciardone:2016ddg}. 
With respect to PNGs, the nonlocal $R^2$-like inflation presents a new target to our future CMB and Large Scale Structure observations reaching $f_{\rm NL} \sim O(1)$ or slightly more \cite{Meerburg:2019qqi,Karagiannis:2018jdt,Castorina:2018zfk,Munoz:2015eqa,Floss:2022grj,Book:2011dz}. 

We organize the chapter as follows. In Sec.~\ref{Sec:formF} we review the NLQG framework and discuss the ghost-free form-factors in maximally symmetric spacetimes (MSS) \cite{Biswas:2016egy,Koshelev:2016xqb,Koshelev:2017tvv,Koshelev:2017ebj}. We highlight that the form factors of de Sitter (dS) background do depend on the background curvature. We present the versions of NLQG that are suitable for realizing exactly the $R^2$-like inflation.   Then we discuss how the ghost-free conditions in dS and quasi-dS (inflationary) background differ slightly and how the inflationary framework does require an extension of NLQG with higher curvature nonlocal terms. This essential extension is called GNLQG which is the result of the recent study \cite{Koshelev:2022olc}. In Sec.~\ref{sec:GNLQG} we discuss in detail the construction of GNLQG within the scope of embedding $R^2$-like inflation. We also present a brief discussion of UV completion aspects of GNLQG. 
 In Sec.~\ref{sec:GNLQG-predictions} we review the predictions of GNLQG and how this theory offers a new understanding to the physics of inflation \cite{Koshelev:2020foq,Koshelev:2022bvg}. We discuss in detail the ways one can observationally probe GNLQG with the future cosmological and gravitational wave probes.
 In Sec.~\ref{sec:lessonGNLQG} we analyze some of the lessons one can learn from the development of GNLQG and discuss the impacts on the popular EFT understanding of inflation.  We highlight how one can end up with some features in EFT which cannot be present in a fundamental UV complete framework. In Sec.~\ref{sec:concout} we conclude with key takeaways from $R^2$-like inflation in GNLQG and further future directions which we can explore on our way of reaching a consistent quantum gravity.  

Throughout the chapter we work with a mostly positive metric signature $(-,+.+,+)$, we set $\hbar=c=1$ and the reduced Planck mass $M_p=\frac{1}{\sqrt{8\pi G}}$. We use overbars to denote background quantities. 

\section{The form factors of quadratic curvature nonlocal gravity in flat and curved backgrounds} 
\label{Sec:formF}

The action of NLQG which is an analytic infinite derivative (AID) extension of Stelle gravity \eqref{Stelleact} is given by 
\begin{equation}
	S^{\rm Nonlocal}_q = \frac{1}{2}\int d^4x\sqrt{-g}\, \Bigg[ M_p^2R+R\Fc_R\LF \square_s \RF R+
	W_{\mu\nu\rho\sigma}\Fc_C\LF \square_s \RF W^{\mu\nu\rho\sigma} \Bigg]
	\label{AIDG}
\end{equation}
where $\square_s=\frac{\square}{\Mc_s^2}$ with $\Mc_s\ll M_p$ being the scale of nonlocality,  and the form factors $\Fc_R\LF \square_s \RF $ and $\Fc_C\LF \square_s \RF$ are the AID operators which can be Taylor expanded as 
\begin{equation}
\Fc_R\LF \square_s \RF = \sum_{m=0}^{\infty} f_{mR} \square_s^{m}\, ,
\quad \Fc_C\LF \square_s \RF =   \sum_{n=0}^{\infty} f_{nC} \square_s^{n}
\label{bforms}
\end{equation}
At this stage the above gravity theory contains an infinite number of arbitrary parameters given by the infinite set $\LF f_{mR},\, f_{mC} \RF$. However, once we demand the theory to be ghost-free, we significantly reduce the freedom of choice for the form factors  \eqref{bforms}. Then it comes to the question what is the background around which the theory has to be ghost-free and how one can find the structure of the form factors. We can first aim to answer these questions in the most simple background we know which is Minkowski spacetime. We want the theory to be ghost-free around Minkowski not only because it is just the simplest spacetime but also we expect that any curved spacetime must be locally Minkowski. This means that at very short length scales or at very high energy scales (but still much lower than the energy scale of possible breaking of the local Lorentz invariance) we expect the spacetime to be nearly Minkowski. On the other hand, since most of realistic cosmological and astrophysical spacetimes are either asymptotically flat or have bounded curvature at spatial infinity, it is indeed a reasonable assumption to demand the gravity theory described by \eqref{AIDG} to be ghost-free around Minkowski.\footnote{Of course, we exclude the discussion of late time effects in some models of present dark energy leading to an unlimited growth of curvature in future. This is another issue indeed and there are constructions of nonlocal gravity theories with both analytic and non-analytic form factors that effect the IR nature of gravity beyond that of the Einstein gravity with an exact cosmological constant \cite{Koivisto:2008xfa,Deser:2019lmm,Bouche:2023xjw,Dimitrijevic:2019pct,Dimitrijevic:2022fhj}.} To obtain ghost-free form factors the method is to obtain the second order perturbation of the action \eqref{AIDG} around the Minkowski background. To be explicit, we consider the general metric perturbation around a chosen background (in this section our discussion is limited to maximally symmetric spacetimes (MSS) such as Minkowski, dS and Anti-dS) as $g_{\mu\nu} = \bar{g}_{\mu\nu}+ \frac{1}{M_p} h_{\mu\nu}$ and consider the 4-dimensional York decomposition of the fluctuations in terms of scalar, vector and tensor parts as 
\begin{equation}
	h_{\mu\nu} = h^\perp_{\mu\nu}+\bar{\nabla}_\mu A_\nu +\bar{\nabla}_\nu A_\mu + \LF \bar{\nabla}_\mu\bar{\nabla}_{\nu}- \bar{g}_{\mu\nu}\bar{\square} \RF B + \frac{1}{4} \bar{g}_{\mu\nu} h  
	\label{metric-decom}
\end{equation}
where $h^\perp_{\mu\nu}$ is the transverse and traceless tensor, $A_\mu$ being the transverse vector, $B,\, h $ are the scalars \cite{Biswas:2016egy,SravanKumar:2019eqt}. Around Minkowski the vector part does not contribute to the second order action and the scalars $B,\,h$ combine into an effective scalar $\varphi = \bar{\square} B-h$. As a result, we obtain the second order perturbed action as 
\begin{equation}
	\begin{aligned}
		\delta^{(2)}S_{q}^{\rm Nonlocal} = \frac{1}{2} \int d^4x\sqrt{-\bar{g}} \LT -\frac{3M_p^2}{32}\varphi \square \Oc_0 \varphi + \frac{M_p^2}{4}h^{\perp}_{\mu\nu} \square\Oc_2 h^{\perp \mu\nu} \RT 
	\end{aligned}
\end{equation}
where the differential operators $\Oc_1$ and $\Oc_2$ take the form \cite{Biswas:2013kla,Biswas:2016egy} 
\begin{equation}
\begin{aligned}
\Oc_0 & =  \LF 1-\frac{6}{M_p^2}\square \Fc_R\LF \square \RF\RF \, ,\\ 
\Oc_2 &=  \LF 1+\frac{2}{M_p^2}\square\Fc_C \LF \square_s \RF \RF \, .
\end{aligned}
\end{equation}
Following the Weierstrass factorization theorem,  if the kinetic differential operators above take  the exponential form, we do not introduce any additional degrees of freedom and as such we completely avoid ghost modes:
\begin{equation}
	\Oc_0 = \LF 1-\frac{\square}{M^2} \RF^p e^{H_0\LF \square_s \RF},\quad \Oc_2 = e^{H_2\LF \square_s \RF}
	\label{O0O2}
\end{equation}
where $H_0\LF \square_s \RF$ and $H_2\LF \square_s\RF$ are the entire functions. And $p=1$ corresponds to the formulation of theory with an additional propagating scalar (called \say{scalaron}) while $p=0$ is the case without any propagating scalar. 
 
Following \eqref{O0O2}, the form factors take the following structure
\begin{equation}
	\begin{aligned}
		\Fc_R\LF \square_s \RF & = \frac{M_p^2}{6} \frac{1-\LF 1-\frac{\square}{M^2}\RF^p e^{H_0\LF \square_s \RF}}{\square} \, , \\  \Fc_C\LF \square_s \RF & = \frac{M_p^2}{2} \frac{e^{H_2\LF \square_s \RF}-1}{\square} \, .
	\end{aligned}
\label{formMin}
	\end{equation}
With the above form factors, 
 the graviton propagator as a function of the 4-momentum square $p^2$ looks like \cite{Krasnikov:1987yj,Koshelev:2016xqb} 
\begin{equation}
	\Pi\LF p^2 \RF \sim -\frac{P^{(2)}}{p^2e^{H_2\LF -p^2 \RF}} + \frac{P^{(0)}}{2p^2\LF 1+\frac{p^2}{M^2} \RF e^{H_0\LF -p^2 \RF}}
	\label{grprop}
\end{equation}
where $P^{(2)},\,P^{(0)}$ are spin projection operators \cite{Biswas:2013kla}. If we require the Newtonian potentials $\Phi,\,\Psi$ of the theory \eqref{AIDG} to be equal around Minkowski, we arrive at the condition \cite{Biswas:2011ar}
\begin{equation}
	H_2\LF \square_s \RF = H_0\LF \square_s \RF
\end{equation} 
which is also known to give a non-singular solution for $\Phi$ \cite{Biswas:2011ar}. It is often speculated that because of the non-singular Newtonian potential, the theory might potentially avoid black hole singularities \cite{Buoninfante:2022ild}. 

Let us now understand if NLQG \eqref{AIDG} including the cosmological constant term
\begin{equation}
	S^{\rm Nonlocal}_{q\Lambda} = \frac{1}{2}\int d^4x\sqrt{-g}\,\Bigg[M_p^2R+R\Fc_R\LF \square_s \RF R+
	W_{\mu\nu\rho\sigma}\Fc_C\LF \square_s \RF W^{\mu\nu\rho\sigma}- \Lambda\Bigg]
	\label{AIDGL}
\end{equation}
is ghost-free. Again, we expect that it entirely depends on the structure of the form factors $\Fc_R\LF \square_s \RF$ and $\Fc_C\LF \square_s\RF$ which can be determined by computing second order action for \eqref{AIDGL} around the MSS background  
\begin{equation}
	\bar{R} = 4\Lambda,\quad  \bar{R}_{\mu\nu} = \frac{\bar{R}}{4}\bar{g}_{\mu\nu},\quad \bar{R}^{\mu}_{\nu\rho\sigma}  =\frac{\bar{R}}{12}\LF \delta^\mu_\rho \bar{g}_{\nu\sigma} -\delta^\mu_\sigma \bar{g}_{\nu\rho}  \RF \, .
\end{equation}
With the York decomposition of the metric fluctuation \eqref{metric-decom}, 
we obtain the second order action of \eqref{AIDGL} as 
\begin{equation}
	\delta^{(2)}S_{q\Lambda}^{\rm Nonlocal} = \frac{1}{2} \int d^4x\sqrt{-\bar{g}} \Bigg[ -\frac{3M_p^2}{32}\varphi   \Oc_{0\Lambda} \LF \bar{\square} +\frac{\bar{R}}{3}  \RF\varphi  + \frac{M_p^2}{4}h^{\perp}_{\mu\nu} \Oc_{2\Lambda} \LF \bar{\square}-\frac{4\Lambda}{6} \RF h^{\perp \mu\nu}\Bigg] 
\end{equation}
where 
\begin{equation}
	\begin{aligned}
\Oc_\Lambda & = 	 \LF 1-\frac{\bar{\square}}{M^2}  \RF^p e^{H_{0\Lambda}\LF \bar{\square}_s,\,\Lambda_s\RF} = 1+ f_{0R} \frac{8\Lambda}{M_p^4} -\frac{2}{M_p^2} \LF 3\bar{\square}+ 4\Lambda \RF \Fc_R\LF \bar{\square}_s \RF\, ,\\ 
\Oc_{2\Lambda} & = e^{H_{2\Lambda}\LF \bar{\square}_s,\,\Lambda_s \RF} = 1+\frac{8\Lambda}{M_p^4}f_{0R}+\frac{2}{M_p^2}\LF \bar{\square}-\frac{\Lambda}{3} \RF \Fc_C\LF \bar{\square}_s+\frac{4\Lambda_s}{3} \RF 
\end{aligned}
\label{formLAID}
\end{equation}
where $\Lambda_s = \frac{\Lambda}{\Mc_s^2}$. 
Here $p=0,\,1$ corresponds to having additional no scalaron and a scalaron degree of freedom respectively. From \eqref{formLAID} we can work out the structure of form factors. To have them analytic, one must appropriately fix the entire functions $H_{0\Lambda},\, H_{2\Lambda}$ such that in the limit $\Lambda\to 0$ we should get back the form factors of Minkowski spacetime \eqref{formMin}. The main lesson we can learn from \eqref{formLAID} is that if we demand the theory to be ghost-free around dS or Anti-dS, we are forced to consider form factors that not only depend on the d'Alembertian operator but they should also be a function of $\Lambda$ in the appropriate form. On the other hand, what we learn here is that the form factors are background dependent and eventually the nonlocal gravity theory we started with \eqref{AIDG} is background dependent as well.  If we want the theory to be background independent, we are somehow forced to consider higher curvature terms. Indeed around curved spacetime, the cubic, quartic and even higher order curvature terms become relevant and unavoidable. This is not something surprising and here is an intuitive explanation. As we can see, \eqref{AIDG} is AID extension of Stelle gravity \eqref{Stelleact}. But logically if we add $W_{\mu\nu\rho\sigma}\square_s W^{\mu\nu\rho\sigma}$, we should equally add $RW_{\mu\nu\rho\sigma} W^{\mu\nu\rho\sigma}$ because both are 6th order derivative terms. The same logic applies to any Nth order term and we cannot omit one term in favor of other. This means the construction of the action \eqref{AIDG} is incomplete and surely we must add more terms to it for consistency and to construct a theory that is ghost-free around curved spacetime. In the several studies in the past, NLQG \eqref{AIDG} is claimed to be the theory that is super-renormalizable and a potential candidate to be UV complete \cite{Modesto:2011kw,Modesto:2014lga,Calcagni:2014vxa,BasiBeneito:2022wux}, but all that is around Minkowski. For any practical application we need a ghost-free theory in curved spacetime and as we saw in the simplest example of dS and AdS spacetimes, we must add higher curvature nonlocal terms. To ascertain the same fact, we are going to further explain this in the context of inflationary scenario from \eqref{AIDG}. Since the action \eqref{AIDG} has infinite derivatives, one thinks that it is impossible to solve the equations of motion (see \cite{Koshelev:2016xqb,Koshelev:2022bvg}). However, due to the structure of the action \eqref{AIDG}, we can solve the equations of motion by proposing an ansatz that leads to recursive relations with the action of d'Alembertian operation on curvature quantities. In the context of cosmological Friedmann-Lemaître-Robertson-Walker (FLRW) backgrounds, we do not have to worry about any contributions to the equations of motion coming from the variation of nonlocal Weyl square term in \eqref{AIDG}. This simplifies significantly our quest for solving equations of motion, and by using the simple eigenvalue equation 
\begin{equation}
	\square R = M^2R
	\label{Staroan}
\end{equation}
we can solve completely equations of motion of \eqref{AIDG} for FLRW backgrounds with the following simple conditions on the form factor $\Fc_R\LF \square_s \RF$ evaluated at $\square = M^2$:
\begin{equation}
	\Fc_R\LF \frac{M^2}{\Mc_s^2} \RF = f_0=  \frac{M_p^2}{6M^2}\,,\quad \Fc^{\dagger}_R\LF \frac{M^2}{\Mc_s^2} \RF = 0 
	\label{condiq}
\end{equation}
where $^\dagger$ denotes derivative with respect to the argument. If we consider the form factor \eqref{formMin}, then \eqref{condiq} exactly imply $H_0(0)=0$ and $p=1$. Clearly this means we need the NLQG \eqref{AIDG} with a propagating scalaron degree of freedom to have \eqref{Staroan} as a background FLRW solution. As a matter of fact, \eqref{Staroan} is exactly the trace equation of the local $R+R^2$ gravity \eqref{Staroact} which solution just provides the Starobinsky inflationary model \cite{Starobinsky:1980te,Koshelev:2016xqb}.  If we require inflationary scenario to be stable against perturbations, we have to require the form factors to be ghost-free. Again to determine this, we have to compute the second order action for \eqref{AIDG}, but in this case it is convenient to define the metric fluctuations in terms of 1+3 decomposition as 
\begin{equation}
ds^2 = a^2( \tau)\LF -\LF 1+2\Phi \RF d\tau^2+ \LF \LF 1-2\Psi\RF \delta_{ij}+h_{ij} \RF  dx^idx^j   \RF\,, 
\end{equation}
where $\Phi$ and $\Psi$ are the Bardeen potentials and $h_{ij}$ is a transverse and traceless tensor. The two Bardeen potentials are constrained during inflation as
\begin{equation}
	\Bigg[f_0\bar{R}_{\rm dS}+ \LF \bar{\square}-\frac{\bar{R}_{\rm dS}}{6} \RF \Fc_C\LF \bar{\square}_s+\frac{\bar{R}_{\rm dS}}{2\Mc_s^2} \RF \Bigg] \LF \Phi+\Psi \RF =0\,.
	\label{phipsi}
\end{equation}
We derive \eqref{phipsi} after linearizing equations of motion of \eqref{AIDG} around the inflationary background satisfying \eqref{Staroan} and then applying the quasi-dS (slow-roll) approximation, i.e., $\bar{R}_{\rm dS}\gg 3M^2$ and slowly changing during inflation (here $\bar{R}_{\rm dS}$ is the value of Ricci scalar in this epoch) \cite{Koshelev:2016xqb}. 
Curiously in \eqref{phipsi} we can notice that the form factor $\Fc_C\LF \square_s \RF$ is involved. In the case of local $R^2$ inflation we know that $\Phi+\Psi\approx 0$ during inflation \cite{Starobinsky:1981zc,Mukhanov:1990me}, but to have the same thing to hold in the context of NLQG \eqref{AIDG}, we must need the operator acting on $\Phi+\Psi$ in \eqref{phipsi} to be exponent of an entire function and at the same time we need to maintain the analyticity property of the form factor $\Fc_C\LF \square \RF$. This implies
\begin{equation}
	\Fc_C \LF \square_s \RF = f_0\bar{R}_{\rm dS} \frac{e^{\gamma_T\LF \square_s-\frac{\bar{R}_{\rm dS}}{3\Mc_s^2} \RF}-1}{\LF \bar{\square}_{\rm dS} -\frac{2\bar{R}_{\rm dS}}{3\Mc_s^2}\RF} \, .
	\label{FCNLQG}
\end{equation}
We can see that the form factor \eqref{FCNLQG} contains the dependence on the background again similar to the case of exact dS \eqref{formLAID}, but here we have the dependence through $\bar{R}_{\rm dS}$ which is treated to be nearly constant but in practice Ricci scalar is not exactly constant but rather slowly varying according to \eqref{Staroan}. 
This means the action \eqref{AIDG} cannot be the fundamental action that describes inflation but rather it must be an effective version of some more fundamental theory of nonlocal gravity. Furthermore, if we compute the second order action of \eqref{AIDG} for tensor perturbation with the form factor \eqref{FCNLQG}, we obtain 
\begin{equation}
	\delta_{(t)}^{(2)} S_{q}^{\rm Non-local} = \int d^4x \sqrt{-g} \Bigg[h_{ij} \, e^{\gamma_T\LF \frac{\bar{	\square}_{\rm dS}}{\Mc_s^2} -\frac{\bar{ R}_{\rm dS}}{3\Mc_s^2} \RF}\LF\bar{	\square}_{\rm dS} -\frac{\bar{R}_{\rm dS}}{6} \RF h^{ij}\Bigg] \, .
	\label{s2h}
\end{equation} 
The above action \eqref{s2h} exactly tell us that NLQG \eqref{FCNLQG} is indeed the ghost-free form factor in the context of inflationary background. 
Substituting \eqref{FCNLQG} in the action \eqref{AIDG}, we can notice terms like 
\begin{equation}
\frac{\bar{R}_{\rm dS}}{\Mc_s^2} W_{\mu\nu\rho\sigma} W^{\mu\nu\rho\sigma}, \frac{\bar{R}_{\rm dS}}{\Mc_s^2} W_{\mu\nu\rho\sigma}\square_s  W^{\mu\nu\rho\sigma}, \LF\frac{\bar{R}_{\rm dS}}{\Mc_s^2}\RF^2 W_{\mu\nu\rho\sigma}\square_s  W^{\mu\nu\rho\sigma}, \frac{\bar{R}_{\rm dS} }{\Mc_s^2}W_{\mu\nu\rho\sigma}\square_s^2 W^{\mu\nu\rho\sigma} \cdots
\end{equation}
which again clearly indicate the necessity of introducing higher order curvature terms, otherwise we end up with an action which has specific background curvature dependence. Another indication why  \eqref{FCNLQG} in \eqref{AIDG} is not a consistent picture is because such a theory would possibly (if $\bar{R}_{dS}\gtrsim \Mc_s^2$) lead to larger production of scalarons decaying into gravitons after the end of inflation according to recent study \cite{Koshelev:2022wqj} (note that in the local $R+R^2$ gravity, the scalaron decay into two gravitons is suppressed \cite{Starobinsky:1981zc}).  This is because if $\bar{R}_{dS}$ is treated as a fixed constant, its effect will be there even after inflation ends. 
In the next section we present a consistent extension of the NLQG \eqref{AIDG} with the higher curvature nonlocal terms where we can see that the form factor \eqref{FCNLQG} emerges naturally at the linearized level in the quasi-dS approximation. 

Let us now discuss scalar perturbations. Considering $\Phi+\Psi \approx 0 $ during inflation, 
 the second order action of \eqref{NAID} for the scalar perturbations become \cite{Koshelev:2017tvv}
\begin{equation}
	\delta_{(s)}^{(2)}S = \frac{1}{2f_0\bar{R}_{\rm dS}} \int d^4x\sqrt{-\bar{g}} \Upsilon\frac{\Wc\LF\bar{\square}_s\RF}{\Fc_R\LF \bar{\square}_s \RF}\LF \bar{\square}_{\rm dS} -M^2 \RF \Upsilon 
	\label{Upsiloneq}
\end{equation}
where $\Upsilon = 2f_0\bar{R}_{\rm dS}\Psi$ is the canonical variable, and it is related to the curvature perturbation as $\Upsilon\approx -2\epsilon f_0\bar{R}_{\rm dS}\Rc$. The operator $\Wc\LF \square_s \RF$ is given by
\begin{equation}
	\Wc\LF \square_s \RF = 3\Fc_R\LF \square_s \RF + \LF \bar{R}_{\rm dS}+3M^2 \RF \frac{\Fc_R\LF \square_s \RF-f_0}{\square-M^2}\,.
	\label{opW}
\end{equation}
From \eqref{Upsiloneq} we can verify that the kinetic term of $\Upsilon$ has one real zero corresponding to  $\bar{\square}_{\rm dS} = M^2$. This means there is one propagating degree of freedom  (i.e., scalaron). 
If any other degrees of freedom exist, they must arise from zeros of the operator $\Wc\LF \square_s \RF$ \eqref{opW}.  Considering $\Wc\LF \square_s \RF =3f_0 e^{\gamma_0\LF \square_s+\frac{\bar{R}_{\rm dS}}{3\Mc_s^2} \RF} $, where $\gamma_0$ is an entire function of $\square_s+\frac{\bar{R}_{\rm dS}}{3\Mc_s^2}$, 
we get no zeros in the entire complex plane. But the consequence is that the form-factor $\Fc_R\LF \square_s \RF$ dependence on the background quantity $\bar{R}_{\rm dS}$ is unavoidable \cite{Craps:2014wga,Koshelev:2020foq}:
	\begin{equation}
		\Fc_R\LF \square_s\RF \equiv  \Fc_R\LF\square_s,\,\bar{R}_{\rm dS}\RF =  f_0\frac{e^{\gamma_0\LF \square_s+\frac{\bar{R}_{\rm dS}}{3\Mc_s^2}\RF} \LF \square_s-\frac{M^2}{\Mc_s^2} \RF+\LF \bar{R}_{\rm dS}+3M^2 \RF }{3\square+\bar{R}_{\rm dS}} \, .
		\label{formds}
	\end{equation}
This choice of form-factor \eqref{formds} again leads to background dependence in the action \eqref{NAID} similar to \eqref{formLAID} and \eqref{FCNLQG} through the factor $\bar{R}_{\rm dS}$. But if we take \eqref{formMin} instead, we completely avoid any background curvature dependence and also we do not have to worry about the Minkowski limit. Calculating $\Wc\LF \square_s \RF$ for \eqref{formMin}, we obtain 
\begin{equation}
	\Wc\LF \square_s \RF = -f_0\bar{R}_{\rm dS} \LF \frac{1-e^{\gamma_S\LF \square_s \RF}}{\square} \RF + 3f_0 e^{\gamma_S\LF \square_s \RF} \, .
	\label{WFMin}
\end{equation}
The zeros of $\Wc\LF \square_s \RF$ can be obtained by solving
\begin{equation}
	\Wc\LF \frac{Z}{\Mc_s^2} \RF = 0 \implies \bar{R}_{\rm dS}\LF \frac{1-e^{\gamma_S\LF  \frac{Z}{\Mc_s^2} \RF}}{Z} \RF = 3e^{\gamma_S\LF  \frac{Z}{\Mc_s^2} \RF} \,.
	\label{characeq}
\end{equation}
To solve \eqref{characeq}, let us consider the following entire function for simplicity:
\begin{equation}
	\gamma_S\LF \square_s \RF = \alpha_1 \square_s\LF \square_s-\frac{M^2}{\Mc_s^2} \RF\,. \label{chformFR}
\end{equation}
Substituting \eqref{chformFR} into \eqref{characeq}, we get no real solutions but instead complex conjugate ones:
\begin{equation}
	\begin{aligned}
		Z = &\, \frac{M^2}{2}+\frac{\Mc_s^2}{2}\LT \LF 2\pi+4q\pi\RF^2+\LF\frac{M}{\Mc_s}\RF^8 \RT^{1/4}\Bigg\{\cos\LT \frac{1}{2}{\rm Arg}\LT 4\pi i\LF q+\frac{1}{2}\RF+\frac{M^4}{\Mc_s^4} \RT \RT \\& + i \sin\LT \frac{1}{2}{\rm Arg}\LT 4\pi i\LF q+\frac{1}{2}\RF+\frac{M^4}{\Mc_s^4} \RT\RT\Bigg\}\\ 
		\approx &\,\Bigg\vert_{M^2\ll \Mc_s^2}  \pm \Mc_s^2 \sqrt{q+\frac{1}{2}}\LF 1 \pm i\RF
	\end{aligned}
	\label{ccstates}
\end{equation}
where $q\geq 1$ is a positive integer. It was known from works based on string field theory models \cite{Koshelev:2007fi,Koshelev:2009ty,Koshelev:2010bf,Arefeva:2008zru} that complex conjugate poles give (classical) degrees of freedom for a coupled system of scalar fields having both positive and negative kinetic terms in equal numbers. 
The question is whether these degrees of freedom are physical and contribute to inflationary correlators. 
Non-local scalar field theories with infinitely many complex conjugate poles are studied in   \cite{Buoninfante:2018lnh,Buoninfante:2020ctr} with the following type of Lagrangians
\begin{equation}
	\Lc_{\phi} = \frac{1}{2}\phi \LF e^{\gamma_\phi\LF \square_s \RF}-1 \RF \phi -V(\phi) 
	\label{scLag}
\end{equation}
where $\gamma_\phi$ being an arbitrary entire function. It was found that the optical theorem both at the tree-level \cite{Buoninfante:2018lnh,Buoninfante:2020ctr} and the one-loop level \cite{Luca} is satisfied despite the presence of complex conjugate poles. This is due to the exact cancellation of contributions from degrees of freedom corresponding to complex conjugate poles. 
Similar to nonlocal theories like \eqref{scLag}, complex conjugate poles also appear in the context of Lee-Wick theories where one can project away the states using new quantum field theory prescriptions. As several investigations suggest \cite{Modesto:2015ozb,Modesto:2016ofr,Anselmi:2017lia,Anselmi:2017ygm,Anselmi:2018kgz,Anselmi:2020lpp,Anselmi:2021hab,Anselmi:2022toe,Liu:2022gun,Frasca:2022gdz}, we may disregard these states as unphysical.\footnote{Alternatively, the imaginary part of the pole can be made small enough (by a contrived choice of entire function) to avoid any classical instabilities \cite{Koshelev:2021orf}. But such a choice of entire function necessarily should depend on the background value of $\bar{R}_{\rm dS}$.} 

Several of these studies are about nonlocal scalar field theories in Minkowski spacetime. In the context of gravitational action action \eqref{AIDG} with \eqref{formMin}, we have no degrees of freedom corresponding to complex conjugate poles around Minkowski spacetime. We can also notice that in the limit $\bar{R}_{\rm dS}\to 0$, we have $\Wc\LF \square_s \RF\to 3f_0 e^{\gamma_S\LF  \square_s\RF}$ (see \eqref{WFMin}) that confirm there exists only scalaron in addition to massless graviton. It is justifiable to use the form factor \eqref{formMin} in the context of inflation because when we quantize inflationary fluctuations, we impose adiabatic vacuum initial conditions for them deep inside the Hubble radius $k\gg aH$ where $H=\frac{\dot a}{a}$ is the Hubble parameter of a spatially flat FLRW model~\cite{Mukhanov:1990me,Koshelev:2020foq}. Therefore, in the limit $k\gg aH$, we can ignore all the complex conjugate modes \eqref{ccstates} by setting initial conditions for them to be zero that implies inflationary correlations are only sourced by scalaron and massless graviton modes. 
 
In summary, we learned that the NLQG action \eqref{AIDG} is incomplete if we consider curved spacetime. This straightforwardly leads us to formulate a consistent quantum gravity theory that go beyond \eqref{AIDG}. Secondly, applying the framework of $R^2$-like inflation we have learned that we must fully avoid any background dependence of the form factors that we can only do by considering higher curvature nonlocal terms. The third point is if we have constructed a ghost-free nonlocal gravity theory for some choice of form factors in the Minkowski spacetime, we can expect appearance of complex conjugate pairs as poles in the propagators for perturbed modes. Due to the fact that these modes do not exist in the local Minkowski limit, we might set their initial conditions to zero but understanding them requires further development of quantum field theory in curved spacetime.\footnote{As a side remark we would like to point out that quantum field theory of nonlocal theories around Minkowski is another subject of investigation because of possible causality violation around scales shorter than nonlocality scale and issues with the Wick rotation. However, it was noted that microcausality violation may not be a problem because we can never probe that regime \cite{Buoninfante:2018mre,Briscese:2019twl}, and also inflation happens to be at length scales larger than the nonlocality scale \cite{Koshelev:2016xqb}. The problem of Wick rotation in nonlocal field theory has also been successfully addressed with a new prescription of contour integration (see \cite{Buoninfante:2022krn,Briscese:2018oyx,Koshelev:2021orf} and the references therein.)}

\section{Generalized nonlocal gravity and $R^2$-like inflation}

\label{sec:GNLQG}

As we discussed in the previous sections, one needs to go beyond the \eqref{AIDG} to have a consistent ghost-free theory for the dynamical backgrounds such as inflation. In this section, we discuss the generalized nonlocal gravity which is compatible with $R^2$-like inflation as a background solution \cite{Koshelev:2022olc}. Note that local $R^2$-inflation is an important guiding principle to build a consistent quantum theory of gravity.  The action for generalized nonlocal gravity is 
\begin{equation}
	\begin{aligned}
		S_H^{\rm Non-local} 
		= 		 &\,\frac{1}{2}\int d^4x\sqrt{-g}\, \Bigg(M_{p}^2R +
		\Bigg[ R\Fc_R\LF \square_s \RF R   +  \LF \frac{M_p^2}{2\Mc_s^2}+ f_0 R_s \RF W_{\mu\nu\rho\sigma}{\mathcal{F}}_{W}\left(\square_{s},\, R_s \right)W^{\mu\nu\rho\sigma}  \\
		&+ \frac{f_0\lambda_c}{\Mc_s^2}\Lc_1\LF \square_{s} \RF R\, \Lc_2\LF \square_{s} \RF R\, \Lc_3\LF  \square_{s}\RF R \\
		& +\frac{f_0\lambda_R}{\Mc_s^2}\Dc_1\LF \square_s \RF R\Dc_2\LF \square_s\RF W^{\mu\nu\gamma\lambda}\Dc_3\LF \square_s \RF W_{\mu\nu\gamma\lambda}
		\\
		& +\frac{f_0\lambda_W}{\Mc_s^2}\Cc_1\LF \square_s \RF W_{\mu\nu\rho\sigma}\Cc_2\LF \square_s\RF W^{\mu\nu\gamma\lambda}\Cc_3\LF \square_s \RF W_{\gamma\lambda}^{\quad\rho\sigma}\Bigg]+\cdots\Bigg)
		\label{NAID}
	\end{aligned}
\end{equation}
which is a significant extension of \eqref{AIDG} in the scope of $R^2$-like inflation. 
Here $R_s= \frac{R}{\Mc_s^2}$ and $\cdots$ represent higher order curvature terms which are only relevant for 4-point cosmological correlations and beyond. In this chapter, we restrict to results of up to 3-point inflationary correlations. In \eqref{NAID} we write all possible nonlocal terms involving Ricci scalar and Weyl tensor. Let us look into these in more detail. The form factors in the first line of \eqref{NAID} can be fixed as 
\begin{equation}
	\begin{aligned}
		\Fc_R & = f_0 M^2 \frac{1-\LF 1-\frac{\square}{M^2} \RF e^{\gamma_S\LF \square_s \RF}}{\square} \, ,\\
		\Fc_W\LF  \square_s,\, R_s\RF  &  = 
	\frac{e^{\gamma_T
			\LF \square_s-\frac{2}{3} R_s \RF}-1}{\square_s-\frac{2}{3} R_s}
		\label{FW}
		\end{aligned}
\end{equation}
where $\gamma_S\LF \square_s \RF$ and $\gamma_T\LF \square_s \RF$ are the entire functions which can be finite degree polynomials or functions of polynomials, so that the theory has only a finite number of free parameters. 
 With a careful observation, we can deduce that \eqref{FW} very much looks like \eqref{FCNLQG} except that this form factor is the function of Ricci scalar rather than of a particular value of it like in \eqref{FCNLQG}. As is easy to see from \eqref{NAID}, our requirement  of having $R^2$-like inflation demands
\begin{equation}
\gamma_S\LF \frac{M^2}{\Mc_s^2} \RF =	\Lc_i\LF \frac{M^2}{\Mc_s^2} \RF  = \Dc_1\LF \frac{M^2}{\Mc_s^2} \RF =0\,. 
	\label{LCDC}
\end{equation}
Obviously the 4th line of \eqref{NAID} is irrelevant for inflationary background since the background Weyl tensor is zero in FLRW. 
We omit Ricci tensor terms because inflation is by definition quasi-dS $\bar{R}_{\mu\nu} \approx \frac{\bar{R}}{4}\bar{g}_{\mu\nu}$, thus the terms involving it can be neglected. Of course, Ricci tensor dependent terms might be important for curvature scales exceeding the inflationary ones towards the Planck scale. 

The cubic terms in \eqref{NAID} (i.e, 2nd, 3rd and 4th lines) form all the possible combinations after judicious weeding of total derivatives using the following non-local generalization of curvature identities derived in \cite{Barvinsky:1994cg,Barvinsky:1994hw} 
\begin{equation}
	\begin{aligned}
		R^\sigma_\lambda  W_{\mu\nu\rho\sigma}W^{\mu\nu\rho\lambda} & = \frac{R}{4}W_{\mu\nu\rho\sigma}W^{\mu\nu\rho\sigma}  \\  \implies R^\sigma_\lambda\Oc\LF \square_s \RF W_{\mu\nu\rho\sigma}\Oc\LF \square_s \RF W^{\mu\nu\rho\lambda} & = \frac{R}{4}\Oc\LF \square_s \RF W_{\mu\nu\rho\sigma} \Oc\LF \square_s \RF W^{\mu\nu\rho\sigma}\, .
	\end{aligned}
\end{equation}
We can notice in \eqref{NAID} that we do not have any terms with single covariant derivatives because we can eliminate all those terms  by adding arbitrary number of total derivatives. For example, suppose we have a term of the form $\mathbb{R}\nabla_\mu\mathbb{R}\nabla^\mu \mathbb{R}$ in the action. Then we can eliminate this term by adding a total derivative of the form $\square \LF \mathbb{R}^3\RF$. We can continue this procedure for any arbitrary number of derivatives until we arrive at 
\begin{equation}
	\Oc_1(\square_s)\mathbb{R}\Oc_2(\square_s)\mathbb{R}\Oc_3(\square_s)\mathbb{R}
	\label{cubpar}
\end{equation}
where $\mathbb{R}$ is the curvature quantity. 
Furthermore, the following non-local generalization of Weyl identities \cite{Barvinsky:1994cg,Barvinsky:1994hw}  are required to write down the cubic nonlocal Weyl tensor terms in \eqref{NAID}
\begin{equation}
	\begin{aligned}
		& W_{[\mu\nu}^{\quad\gamma\lambda}\Oc\LF \square_s \RF W_{\gamma\lambda}^{\quad\alpha\beta}\Oc\LF \square_s \RF W_{\alpha]\sigma}^{\quad\mu\nu} = 0 \, ,\\ 
		&	W^{\mu\nu\rho\sigma}\Oc\LF \square_s\RF   W_{\mu\nu\rho\lambda}  = \frac{1}{4} \delta^\sigma_\lambda   W^{\mu\nu\rho\alpha}\Oc\LF \square_s \RF W_{\mu\nu\rho\alpha}\, . 
	\end{aligned}
	\label{weylid}
\end{equation} 
In the first line of \eqref{weylid} we have complete anti-symmetrization over the five indices. 
Furthermore, we can in principle make the form factors $\Cc_i\LF \square_s \RF$ to depend on $R_s$, but we simply drop here this generalization for brevity. 

The action (\ref{NAID}) must be viewed as leading terms in the series expansion in the approximation $R\ll M_P^2$, and we can expect more higher curvature terms that might need to be added to build a full quantum gravity action.
Furthermore, we can express the cubic nonlocal form factors as
\begin{equation}
	\begin{aligned}
		\Lc_i\LF \square_s \RF & =  e^{\ell_i\LF\square_s\RF}-1\, ,\\
			\Dc_i\LF \square_s \RF & =  e^{d_i\LF\square_s\RF}-1 \, ,\\
				\Cc_i\LF \square_s \RF & =  e^{c_i\LF\square_s\RF}-1
	\end{aligned}
	\label{cubicformd}
\end{equation}
where $\ell_{i}\LF \square_s \RF,\,c_{i}\LF \square_s \RF,\,d_{i}\LF \square_s \RF$ are the entire functions which we can assume to be polynomials or functions of polynomials that gives us a finite parameter space. To have a UV completion, it is expected that the cubic nonlocal operators $\Lc_i\LF \frac{p^2}{\Mc_s^2} \RF, \,\Dc_i\LF \square_s \RF,\, \Cc_i\LF \square_s \RF$ do not grow in the limit $p\to \infty$. This is also a needed behavior in order to have a consistent inflationary predictions as well. Because if these operators are highly suppressed in the $p\to \infty$ limit, all the loop contributions are expected to be subdominant and the computation of inflationary perturbations at the linear level gives us consistent predictions for correlators.  

From \eqref{cubicformd} and \eqref{LCDC}  we can write a generic form of entire functions $\gamma_S\LF \square_s \RF$ and $\ell_i\LF \square_s \RF$
\begin{equation}
	\begin{aligned}
		\gamma_S\LF \square_s \RF = & \LF \square_s-\frac{M^2}{\Mc_s^2} \RF P_S\LF \square_s \RF \, , \\ 
		\ell_{i}\LF \square \RF = & \LF \square_s-\frac{M^2}{\Mc_s^2} \RF G_i\LF \square_s \RF  \, ,\\ 
			d_{1}\LF \square \RF = & \LF \square_s-\frac{M^2}{\Mc_s^2} \RF D_1\LF \square_s \RF 
	\end{aligned}
	\label{entchoice}
\end{equation}
where $P_S\LF \square_s \RF,\, G_i\LF \square_s \RF,\,  D_1\LF \square_s \RF$ are the finite degree polynomials. 

Before going to the next section where we discuss inflationary observables of $R^2$-like inflation in GNLQG \eqref{NAID}, let us make here an intuitive remark that the presence of non-local term $RW_{\mu\nu\rho\sigma}\Fc_W\LF \square_s,\, R_s \RF W^{\mu\nu\rho\sigma}$ does not necessarily spoil the power counting renormalizability which was achieved in NLQG \cite{Modesto:2011kw}. 
We can easily verify this by taking the the generalized ghost-free form factor \eqref{FW} of the action \eqref{NAID} in the high energy limit, the N-point graviton vertices arising from $W_{\mu\nu\rho\sigma}\Oc\LF \square_s \RF W^{\mu\nu\rho\sigma}$ dominate over the $N-$point vertices arising from $R^{N-2}W_{\mu\nu\rho\sigma}W^{\mu\nu\rho\sigma}$. 
 Therefore,  the traditional power counting renormalizability around Minkowski background which is studied in~\cite{Koshelev:2016xqb} most likely holds for GNLQG \eqref{NAID} especially if we arrange cubic nonlocal form factors such that they are highly suppressed in the high energy limit. However, still renormalizability might require more additional terms to those present in \eqref{NAID} that is an open problem and requires further investigation.

\section{Predictions of generalized nonlocal $R^2$-like inflation}

\label{sec:GNLQG-predictions}

In the previous section, we established $R^2$-like inflation in GNLQG and in this section we review the predictions of the model with the two point and 3-point inflationary correlations. These results are derived in \cite{Koshelev:2022olc,Koshelev:2022bvg} which we briefly review here. 

The second order perturbation of the action \eqref{NAID} around the inflationary solution follows from $\bar{ \square} \bar{ R}=M^2 \bar R$ can be calculated as
\begin{equation}
	\delta^{(2)}S_H^{\rm Non-local} =   \delta^{(2)}S_{R+R^2}^{\rm local}+\delta^{(2)}S_{R+R^2}^{\rm Non-local} + \delta^{(2)}S_{\mathbb{R}^3}^{\rm Non-local}
	\label{2rdv}
\end{equation}
where $S_{R+R^2}^{\rm local}$ is the local $R^2$ action and
\begin{equation}
	\begin{aligned}
		S_{R+R^2}^{\rm Non-local}  = &\, S_{R^2}^{\rm Non-local} + S_{W^2}^{\rm Non-local} \\  =
		&\, \frac{1}{2}\int d^4x\sqrt{-g} \Bigg\{R \Bigg[\Fc_R\LF\square_s\RF-f_0\Bigg]R\\ &+\LF\frac{M_p^2}{2\Mc_s^2}+ f_0\frac{R}{\Mc_s^2} \RF  W_{\mu\nu\rho\sigma}{\mathcal{F}}_{W}\left(\square_{s},\,\frac{R}{\Mc_{s}^2}\right)W^{\mu\nu\rho\sigma}\Bigg\} 
		\label{nlcl2q}
	\end{aligned}
\end{equation}
and $S_{\mathbb{R}^3}^{\rm Non-local}$ constitute from the last 3-lines in \eqref{NAID} which are cubic in curvature.
Applying the conditions \eqref{LCDC}, we can see that the cubic non-local term has no contribution to the above second order action.  
This is not very surprising and we can understand this the following an analogous simple example with the following action 
	\begin{equation}
		S^{\rm Min}_{R^3} = \frac{1}{2}\int d^4x\sqrt{-g}\,  \LT M_p^2R+ f_0R^2+f_0\Lambda_m^{-2}R^3 \RT\,
		\label{toyex}
	\end{equation}
	where $\Lambda_m$ is some mass scale. From the above we can notice that that the cubic term $R^3$ does not effect the second order action (or the linearized equations of motion) around Minknowski, however, it is obviously not true in other backgrounds. If we study inflation with local higher curvature extension of $R^2$, we do modify inflationary solution unless we assume $R^2$ is the most relevant term during inflation. In fact, slow-roll inflation in local $f(R)$ gravity occurs for the range of $R$ for which the Lagrangian density ${\cal L}$ is close to $R^2$, namely, ${\cal L}=A(R)R^2$ where $A(R)$ is a slowly changing function of $R,~|\frac{d\ln A}{d\ln R}|\ll 1$~\cite{Appleby:2009uf}. In our case the cubic scalar curvature nonlocal terms (i.e.,  the last 3 lines in \eqref{NAID}) are introduced in such a way that we do not get any contributions to the two point correlations which means
\begin{equation}
	\delta^{(2)}S_{\mathbb{R}^3} \Bigg\vert_{\bar{\square} \bar{ R} = M^2\bar{R}} = 0\,.
\end{equation}
However, the cubic nonlocal terms are expected to contribute to the 3rd order perturbation of the action \eqref{NAID} which implies the last 3 lines in \eqref{NAID} effect the primordial non-Gaussianities which we discuss in the next subsection. 
After long computations, we deduce that second order action in \eqref{2rdv} exactly coincides with the result obtained in the context of NLQG \cite{Koshelev:2017tvv} which is not surprising because our construction of GNLQG is an extension of NLQG preserving the already interesting inflationary predictions of $R^2$-like inflation in NLQG  \cite{Koshelev:2016xqb,Koshelev:2017tvv,Koshelev:2020foq}. 
The crucial difference in GNLQG is that we consider a choice of $\Fc_R\LF \square_s \RF,\,\Fc_W\LF \square_s \RF$ that does not explicitly depend on the background curvatures during inflation which was the case of earlier NLQG frameworks \cite{Craps:2014wga,Koshelev:2017tvv,Koshelev:2020foq}. This ascertain that GNLQG is more fundamental action and is more towards the full description of quantum gravity in the nonlocal setup. 

To compute the scalar power power spectrum we start with the second order action for the scalar perturbation given in \eqref{Upsiloneq} which happened to be the same both in NLQG and GNLQG. We use the form factor \eqref{FW} for which we have seen from \eqref{WFMin} that there is only one scalaron and we set the initial conditions of all the modes corresponding to complex conjugate poles. Thus the calculation of scalar power spectrum $\Pc_\Rc$ and spectral index $n_s$ of $R^2$-like inflation in GNLQG are pretty much standard by the use of canonical rescaling of the fields and the final results are \cite{Koshelev:2022olc}
\begin{equation}
	\Pc_{\Rc} (k)\approx \frac{1}{3f_0\bar{R}_{\rm dS}} \frac{H^2}{16\pi^2\epsilon^2}\Bigg\vert_{k=aH}\,,\quad n_s-1 \equiv \frac{d\ln \Pc_{\Rc}}{d\ln k }\Bigg\vert_{k=aH} \approx -\frac{2}{N}\,, 
	\label{pr}
\end{equation}
where $a(t)H(t)$ is estimated at the first Hubble radius crossing during inflation for a given mode wavenumber $k$. 
%$k_\ast=a_\ast H_\ast = 0.05\,{\rm Mpc}^{-1}$ is some pivot scale \cite{Akrami:2018odb} at which the power spectrum is determined as %$\Pc_\Rc \approx 2.1\times 10^{-9}$. 
The above result \eqref{pr} is almost independent of $\Fc_R\LF \square_s\RF$ as far as $M^2\ll \Mc_s^2$, and the dimensionless coefficients of $\Fc_R\LF \square_s\RF$ \eqref{FW} are of the $O(1)$. Moreover, the scalar slope $n_s$ is practically the same as in the Starobinsky inflationary model \eqref{Staroact}~\cite{Mukhanov:1981xt,Starobinsky:1983zz}, see \cite{Koshelev:2017tvv,Koshelev:2020foq} for further details where it is shown that the non-local correction to it is of the order of $O\LF\frac{M^4}{\Mc_s^4}\RF$. 

\begin{figure}
	\centering
	\includegraphics[width=0.7\linewidth]{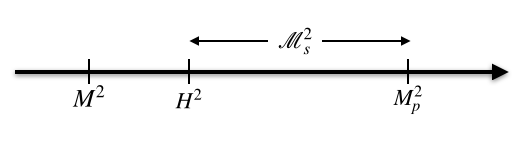}
	\caption{The hierarchy of in the generalized non-local $R^2$-like inflation.}
	\label{fig:scales}
\end{figure}

\begin{figure}[h]
	\centering
	\includegraphics[width=0.6\linewidth]{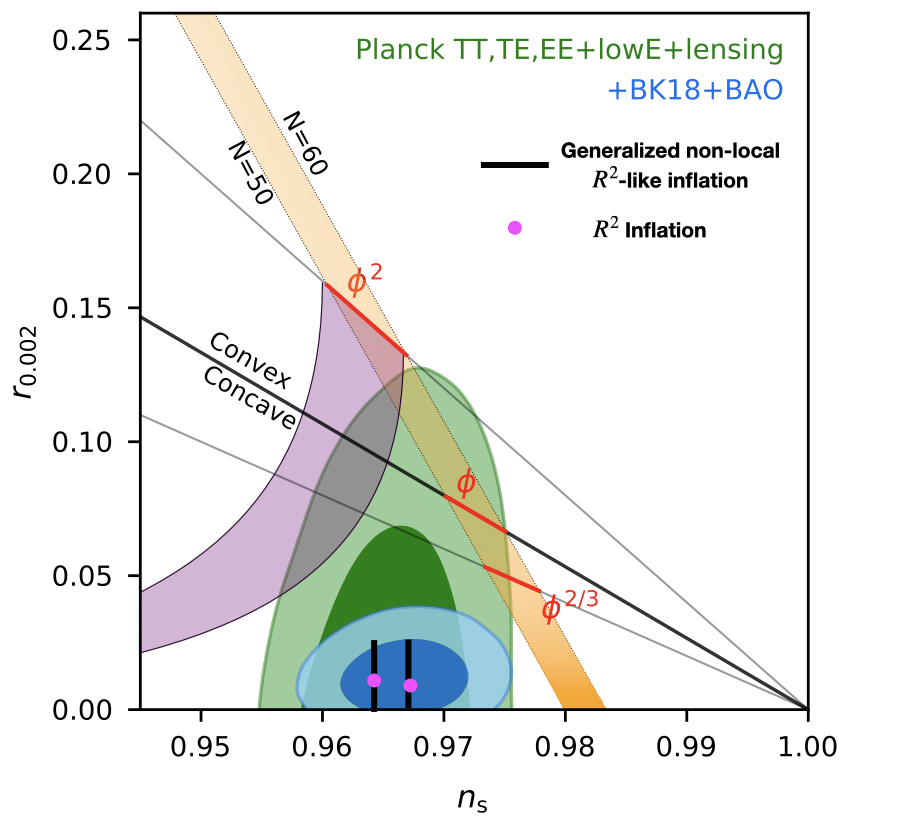}
	\caption{Here we depict the predictions of generalized nonlocal $R^2$-like inflation  with respect to the Planck and BICEP/Keck array analysis (with the ruled out monomial models of inflation and the natural inflation shown in purple colour) which has constrained the tensor-to-scalar ratio as $r<0.036$ and the spectral index as $n_s=0.9649\pm 0.0042$. The black colour vertical line from the left represents $\LF n_s,\,r\RF = \LF 0.964,\, < 0.036 \RF$ for $N=55$, whereas the one on the right represents $\LF n_s,\,r\RF = \LF 0.967,\, < 0.036 \RF$ for $N=60$. }
	\label{nsr}
\end{figure}

In the inflationary high curvature regime $R\gg M^2$, quadratic curvature terms are naturally dominant. Thus, the hierarchy $M^2\ll \Mc_s^2$ is essential to have the generalized nonlocal $R^2$-like inflation to be compatible with Planck data. On the other hand, $H^2$ can be of the order or even larger than $\Mc_s^2$ while still being much less than $M_p^2$, see the hierarchy of scales in Fig.~\ref{fig:scales}. 
 This can be seen 
heuristically by expanding the quadratic Ricci scalar part of the action (\ref{NAID}) as 
\begin{equation}
	S = \int d^4x\sqrt{-g} \LT \frac{M_p^2}{2}R+\frac{M_p^2}{12M^2}R^2+O \LF \frac{M_p^2 R\square R}{M^2\Mc_{s}^2}  \RF+\cdots \RT\,.
\end{equation}
In order to compute the tensor power spectrum we use the second order action for the tensor mode \eqref{s2h} which turned out to be exactly same in both NLQG and GNLQG \cite{Koshelev:2022olc}. To compute tensor power spectrum we do the rescaling of $h_{ij}\to e^{\gamma_T\LF \bar{\square}_{\rm dS}-\frac{\bar{R}_{\rm dS}}{3\Mc_s^2}\RF}h_{ij}$ and bring the action \eqref{s2h} into the local form. Then after performing the canonical quantization of the field by standard methods we will rescale back the result by evaluating the exponent of entire function at the pole $\bar{\square}_{\rm dS} = \frac{\bar{R}_{\rm dS}}{6}$ \cite{Koshelev:2016xqb,Koshelev:2022olc}. 
As a result, we get the inflationary tensor power spectrum of generalized non-local $R^2$ model as
\begin{equation}
	\begin{aligned}
		\Pc_T & = \frac{1}{12\pi^2f_0} \LF 1-3\epsilon \RF e^{-2\gamma_T\LF \frac{-\bar{	R}_{dS}}{6\Mc_{s}^2} \RF} \Bigg\vert_{k=a H}\,,
	\end{aligned}
	\label{pt}
\end{equation}
The tensor-to-scalar ratio from \eqref{pr} and \eqref{pt} can be calculated as 
\begin{equation}
	r = \frac{12}{N^2} e^{-2\gamma_T\LF -\frac{\bar{	R}_{\rm dS}}{6\Mc_s^2} \RF}\Bigg\vert_{k=aH}\,. 
	\label{t2s}
\end{equation}
The tensor spectral index and its running and running of running can be evaluated as
\begin{equation}
	\begin{aligned}
		n_t \equiv \frac{d\ln\Pc_T}{d\ln k }\Bigg\vert_{k=aH} & \approx -\frac{3}{2N^2}-\LF \frac{2}{N}+\frac{1}{N^2} \RF \frac{\bar{	R}_{\rm dS}}{6\Mc_s^2} \gamma_T^{\dagger}\LF -\frac{\bar{	R}_{\rm dS}}{6\Mc_{s}^2} \RF \, \\ 
		\frac{dn_t}{d\ln k} \Bigg\vert_{k=aH} & \approx -\frac{3}{N^3}-\frac{1}{N^3}\frac{\bar{R}_{\rm dS}}{6\Mc_s^2}\gamma_T^{\dagger}\LF -\frac{\bar{	R}_{\rm dS}}{6\Mc_{s}^2} \RF -\frac{1}{18N^2}\frac{\bar{	R}_{\rm dS}^2}{\Mc_{s}^4}  \gamma_T^{\dagger\dagger}\LF -\frac{\bar{	R}_{\rm dS}}{6\Mc_{s}^2} \RF \,\\
		\frac{d^2n_t}{d\ln k^2} \Bigg\vert_{k=aH} & \approx -\frac{9}{N^4}-\frac{1}{3N^4}\frac{\bar{R}_{\rm dS}}{\Mc_s^2}\gamma_T^{\dagger}\LF -\frac{\bar{R}_{\rm dS}}{6\Mc_s^2} \RF -\frac{1}{12N^3}\frac{\bar{R}_{\rm dS}^2}{\Mc_s^4}\gamma_T^{\dagger\dagger}\LF -\frac{\bar{R}_{\rm dS}}{6\Mc_s^2}\RF \\ &\quad -\frac{1}{108N^3}\frac{\bar{R}_{\rm dS}^3}{\Mc_s^6}\gamma_T^{\dagger\dagger\dagger}\LF -\frac{\bar{R}_{\rm dS}}{6\Mc_s^2}\RF
	\end{aligned}
	\label{ntt}
\end{equation}
where $^{\dagger},\,^{\dagger\dagger},\,^{\dagger\dagger\dagger}$ indicate first, second and third derivative with respect to the argument, we used $\frac{d}{d\ln k}\approx -\LF 1+\frac{1}{2N} \RF\frac{d}{dN}$ and $\frac{d\bar{R}_{\rm dS}}{dN}\approx 2\bar{R}_{\rm dS}\epsilon$. 
From \eqref{pt}, \eqref{t2s} and \eqref{ntt} we can conclude the following 
\begin{itemize}
	\item The tensor power spectrum is modified due to the higher curvature nonlocal terms involving Ricci scalar and Weyl tensor in \eqref{NAID}.
	\item Compared to the result in the local $R+R^2$ gravity \eqref{Staroact}~\cite{Starobinsky:1983zz}, the tensor power spectrum contains a strong scale dependence due to the exponential term $e^{-2\gamma_T\LF\frac{-\bar{R}_{\rm dS}}{6\Mc_s^2}\RF}$ where $\bar{R}_{\rm dS}(k)$ depends on wave number $k$. 
	\item With tensor tilt and its running and running of running computed in from \eqref{ntt} we can probe $\gamma_T\LF -\frac{\bar{R}_{\rm dS}}{6\Mc_s^2}\RF$ and reconstruct the form-factor \eqref{FW}  from the future primordial gravitational wave observations \cite{CMB-S4:2020lpa,Calcagni:2020tvw}. 
	\item The single field tensor consistency relation $r=-8n_t$ gets violated solely due to the modification of tensor-power spectrum. 
\end{itemize}
In Fig.~\ref{nsr} we depict the status of $R^2$-like inflation in GNLQG against the latest constraints from Planck+BICEP/Keck analysis where we can see that in this model we can have any tensor-to-scalar ratio $r<0.036$. This has been shown explicitly by the simplest choices of entire function \cite{Koshelev:2022olc}
\begin{equation}
	\gamma_T\LF \square_s-\frac{2R_s}{3} \RF = \beta_1 \LF  \square_s-\frac{2R_s}{3} \RF^2+\beta_2\LF  \square_s-\frac{2R_s}{3} \RF^3+\cdots. 
\end{equation}
where $\cdots$ denote possible higher order terms which can be irrelevant in the context of inflation but such terms could play a role in taming UV divergences. In the next section, we shall come back to these predictions and compare them with the EFT models of inflation. 

\begin{figure}[h]
	\centering
	\includegraphics[width=0.6\linewidth]{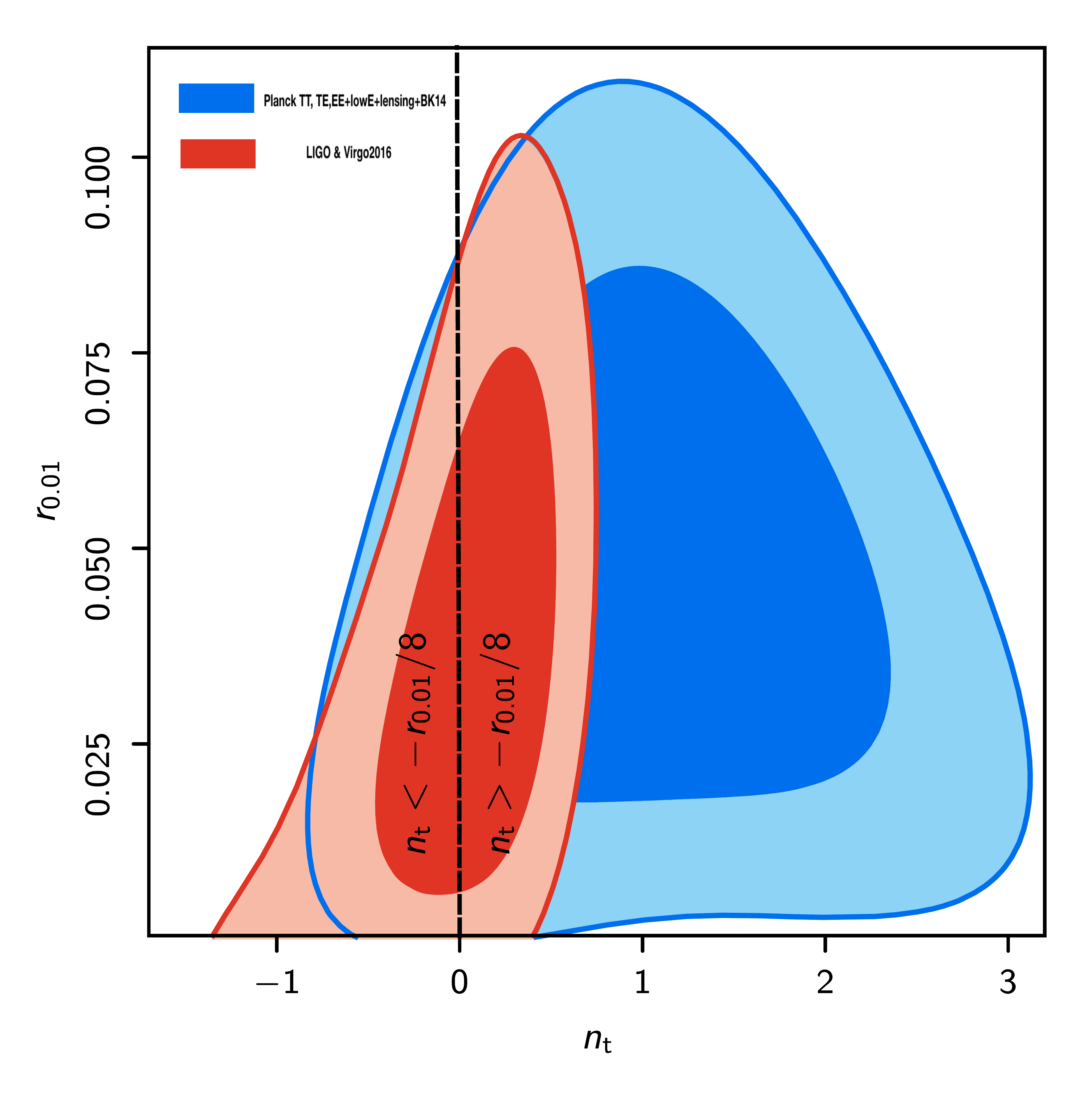}
	\caption{This figure represents $n_t-r$ plane from Planck 2018 data. The predictions of generalized nonlocal $R^2$-like inflation can be anywhere within the plane. If future experiments detect actual values of these quantities, we shall be able to constrain more precisely the form factors and the scale of nonlocality. }
	\label{fig:nt-r}
\end{figure}

In Fig.~\ref{fig:nt-r} we depict the fact that the generalized non-local $R^2$-like inflation can predict both positive and negative values of tensor tilt within the likelihood region from the Planck data. 

\subsection{Primordial (scalar) Non-gaussianities}

PNGs are of enormous importance in the $R^2$-like inflation in GNLQG. Because, when we calculate interaction vertices in nonlocal graivty we most often get the presence of AID operators unlike the local models of inflation \cite{Chen:2010xka}. Therefore, PNGs are the most pertinent observables to find the signature of nonlocalities. In this section we review only the scalar PNGs in GNLQG which are derived in \cite{Koshelev:2022bvg}.

PNGs are calculated with computing 3-point correlations defined by \cite{Maldacena:2002vr,Koshelev:2020foq} 
\begin{equation}
	\langle\Rc\left(\mathbf{k_{1}}\right)\Rc\left(\mathbf{k_{2}}\right)\Rc\left(\mathbf{k_{3}}\right)\rangle  = -i \int_{-\infty}^{\tau_e}   d\tau \langle 0\vert [ \Rc(\tau_{e},\,\mathbf{k_1})\Rc(\tau_{e},\,\mathbf{k_2})\Rc(\tau_{e},\,\mathbf{k_3}),\, H_{int} ] \vert 0 \rangle \,,
	\label{3-point-f}
\end{equation}
where $\boldsymbol{k}_i$ are wave vectors and $H_{int}\approx -\Lc_3$ is the interaction Hamiltonian that is approximately equal to the 3rd order perturbation of the Lagrangian (\ref{NAID}) ($\Lc_3$) within the slow-roll approximation  \cite{Maldacena:2002vr,DeFelice:2011zh} and $\tau_e$ denotes the end of inflation.

The bi-spectrum ($\Bc_\Rc$) is usually defined as
\begin{eqnarray}
	\langle\Rc\left(\mathbf{k_{1}}\right)\Rc\left(\mathbf{k_{2}}\right)\Rc\left(\mathbf{k_{3}}\right)\rangle  =  \left(2\pi\right)^{3}\delta^{3}\left(\mathbf{k_{1}}+\mathbf{k_{2}}+\mathbf{k_{3}}\right)\mathcal{B}_{\Rc}\left(k_{1},k_{2},k_{3}\right)
\end{eqnarray}
where $\vert\boldsymbol{k_i}\vert=k_i$ and 
the non-linear curvature perturbation $\Rc$ is expressed as 
\cite{Komatsu:2001rj,Takahashi:2014bxa}
\begin{equation}
	\Rc = \Rc_g  -\frac{3}{5}f_{NL}\LF \Rc_g^2-\langle \Rc_g \rangle ^2 \RF \,, 
\end{equation} 
where $\Rc_g$ being the Gaussian random field and the $f_{NL}$ is the non-linearity parameter also known as the reduced bi-spectrum \cite{Kenton:2015lxa} that is defined as
\begin{equation}
	f_{NL} = -\frac{5}{6}\frac{A_\Rc\LF k_1,\,k_2,\,k_3 \RF}{\sum_i k_i^3}\,, 
	\label{fnldef}
\end{equation}
where $A_\Rc\LF k_1,\,k_2,\,k_3 \RF$ stands for the redefinition of the bi-spectrum $B_\Rc$: 
\begin{equation}
	B_{\Rc}\left(k_{1},k_{2},k_{3}\right) = 4\pi^4\frac{1}{\prod_i k_i^3} \Pc_{\Rc}^{ 2} A_{\Rc}\LF k_1,\,k_2,\,k_3 \RF \, .
\end{equation}
To calculate $f_{\rm NL}$, we calculate the third order variation of \eqref{NAID} around the background \eqref{Staroan} as
\begin{equation}
	\delta_{(s)}^{(3)}S_H^{\rm Non-local} =   \delta_{(s)}^{(3)}S_{R+R^2}^{\rm local}+\delta_{(s)}^{(3)}S_{R+R^2}^{\rm Non-local} + \delta_{(s)}^{(3)}S_{\mathbb{R}^3}^{\rm Non-local}
	\label{3rdv}
\end{equation}
where 
\begin{equation}
	\begin{aligned}
		S_{R+R^2}^{\rm local}  = & \int d^4x\sqrt{-g}\Bigg[\frac{M_p^2}{2}R+ \frac{f_0}{2} R^2 \Bigg] \\ 
		S_{R+R^2}^{\rm Non-local} = &  \int d^4x\sqrt{-g}\Bigg\{\frac{M_p^2}{2}R+ \frac{1}{2} R\Bigg[\Fc_{R}\LF \square_s \RF -f_0\Bigg] R \Bigg\} \\ 
		S_{R^3}^{\rm Non-local} = &  \int d^4x\sqrt{-g}\Bigg[\Lc_1\LF \square_s\RF R\Lc_2\LF \square_s\RF R\Lc_3\LF \square_s \RF R \Bigg] \\
	\end{aligned}
	\label{listact}
\end{equation}
and the subscript $_{(s)}$ in \eqref{3rdv} denotes the scalar part of the 3-rd order action as we are not discussing tensor PNGs here. 
Using the calculations performed in \cite{Koshelev:2020foq}, since $\Phi+\Psi\approx 0 $ during inflation and the variation $\delta_{(s)}W_{\mu\nu\rho\sigma}\propto \Phi+\Psi$, we can conclude that all the terms involving Weyl tensor do not contribute to the scalar PNGs. In \eqref{NAID}, we can further possibly consider quartic order non-local scalar curvature term:
\begin{equation}
	S^{\rm Non-local}_{R^4} = \frac{f_0\lambda_q}{2\Mc_s^4}\int d^4x\sqrt{-g}\Bigg[ \Lc_4 \LF \square_s \RF R \Lc_1 \LF \square_s \RF R \Lc_2 \LF \square_s \RF R \Lc_2 \LF \square_s \RF R\,  \Bigg]. 
	\label{r4nl}
\end{equation}
Here $\Lc_4\LF \square_s \RF$ is an arbitrary analytic infinite derivative operator. It is easy to deduce that \eqref{r4nl} still admits the inflationary solution \eqref{Staroan}. Applying  $\Lc_i\LF \frac{M^2}{\Mc_s^2} \RF =0$ ($i=1,2,3$), we can conclude that
the second order variation of \eqref{r4nl} around the background satisfying \eqref{mR2in} becomes zero exactly 
\begin{equation}
	\delta^{(2)}S_{R^4}^{\rm Non-local} =0\,, 
\end{equation}
whereas the 3rd order variation of \eqref{r4nl} around \eqref{Staroan} in the leading order de Sitter approximation is 
\begin{equation}
	\delta^{(3)}S_{R^4}^{\rm Non-local} \approx \Lc_4\LF \frac{M^2}{\Mc_s^2} \RF\frac{\lambda_q \bar{R}_{\rm dS}}{\lambda_c\Mc_s^2}  \delta^{(3)}S_{R^3}^{\rm Non-local} \,. 
\end{equation}
Given that
\begin{equation}
	\Lc_4\LF \frac{M^2}{\Mc_s^2} \RF\frac{\lambda_q \bar{R}_{\rm dS}}{\lambda_c\Mc_s^2}  \ll 1\, ,
\end{equation}
we can neglect \eqref{r4nl} for the scalar (3-point) PNGs. However, we can expect to have non-negligible contributions to the trispectrum (or) 4-point correlations due to this term. The same logic can be easily extended to other higher order scalar curvature nonlocal terms as well as for the higher order Weyl curvature nonlocal terms. 

Therefore, we can conclude that nonlocal contributions to the bispectrum arise only from the part of the action \eqref{NAID} which is quadratic and cubic in Ricci scalar. After the long computations we can rewrite \eqref{3rdv} in terms  of curvature perturbation $\Rc$ as 
\begin{equation}
	\begin{aligned}
	\delta_{(s)}^{(3)}S_{H}^{\rm Non-local} = \, & 4\frac{M_p^2}{H^2}\epsilon  \int   d\tau d^3x \Bigg\{B_1 \tau^{-2} \Rc\nabla\Rc\cdot\nabla\Rc+
	B_2 \tau^{-2} \Rc\Rc^{\prime 2}+  B_3 	\tau^{-3} \Rc\Rc\Rc^\prime\\+&
	B_4 \tau^{-1}\Rc^{\prime 3}+ B_5 
	\tau^{-4}\Rc^3+ B_6\tau^{-1}\nabla\Rc\cdot\nabla\Rc\Rc^\prime + B_7 \Rc^\prime \nabla\Rc\cdot\nabla\Rc^\prime  \Bigg\}\,, 
	\end{aligned}
	\label{3rda}
\end{equation}
where $B_1$ to $B_7$ are dimensionless parameters approximated to be constant during inflation which are given by 
\begin{equation}
	\begin{aligned}
	B_1 & = -2\epsilon-\frac{3\epsilon^2}{4} \\ 
B_2 & = 2\epsilon+ \frac{3\epsilon^2}{4} +\frac{16}{3}\epsilon \Tc_{\rm NL} + \frac{8}{3} \epsilon ^3 \frac{ \bar{R}_{\rm dS}^2 }{ \Mc_S^4} \gamma_S^\dagger\LF \frac{\bar{R}_{\rm dS}}{4\Mc_s^2} \RF e^{ \gamma_S\LF \frac{\bar{R}_{\rm dS}}{4\Mc_s^2} \RF  } -\frac{2\lambda_c}{9}\frac{\bar{R}_{\rm dS}}{\Mc_s^2}\epsilon\LF 2\epsilon^2T_{\rm NL}^3+\epsilon T^2_{\rm NL}+ T_{\rm NL}^1 \RF\\
B_3 & = -32\Tc_{\rm NL}-\frac{8\lambda_c}{9}\frac{\bar{R}_{\rm dS}}{\Mc_s^2}\epsilon\LF 2\epsilon T_{\rm NL}^2+T_{\rm NL}^1 \RF \\ 
B_4 & = -2 \Tc_{\rm NL} -\frac{1}{54}\frac{\bar{R}_{\rm dS}}{\Mc_s^2}\epsilon\LF 8\epsilon^3 T_{\rm NL}^4+4\epsilon^2T_{\rm NL}^3+2\epsilon T_{\rm NL}^2+T_{\rm NL}^1 \RF \\ 
B_5 & = -\frac{\epsilon^2}{2}+\frac{32\bar{R}_{\rm dS}}{\Mc_s^2}\epsilon T_{\rm NL}^1\,\\
B_6 & = -2\Tc_{\rm NL}\\
B_7 & =  \frac{16}{3}\epsilon \Tc_{\textrm{NL}}+\frac{8}{3} \epsilon ^3 \frac{ \bar{R}_{\rm dS}^2 }{ \Mc_S^4} \gamma_S^\dagger\LF \frac{\bar{R}_{\rm dS}}{4\Mc_s^2} \RF e^{ \gamma_S\LF \frac{\bar{R}_{\rm dS}}{4\Mc_s^2} \RF  } \, ,  
	\end{aligned} 
\label{intr}
\end{equation}
where  the subscript \say{$_{\rm dS}$} denotes the quantities in quasi-dS approximation. 
And the quantities $\Tc_{\rm NL}, T_{\rm NL}^1,\cdots, T_{\rm NL}^4$ are given by
\begin{equation*}
	\begin{aligned}
			\Tc_{\rm NL} & =  \frac{\bar{R}_{\rm dS}}{M_p^2}\epsilon^3   \LT\Fc_{R}\LF \frac{M^2}{\Mc_{s}^2} +\frac{\bar{ R}_{\rm dS}}{4\Mc_{s}^2}\RF - \Fc_{1} \RT  \approx \frac{\bar{R}_{\rm dS}}{M_p^2}\epsilon^3 \LT  \Fc_{R}\LF \frac{\bar{ R}_{\rm dS}}{4\Mc_{s}^2}\RF - \Fc_{1} +\epsilon\frac{\bar{ R}_{\rm dS}}{8\Mc_{s}^2}\Fc^{(\dagger)}_R\LF \frac{\bar{ R}_{\rm dS}}{4\Mc_{s}^2}\RF \RT
			\\ & \approx \frac{1}{3}\epsilon ^2\left(e^{\gamma_S\LF \frac{\bar{R}_{\rm dS}}{4\Mc_s^2} \RF }-1\right) +\epsilon ^3 \frac{ \bar{R}_{\rm dS}^2 }{12 \Mc_S^4} \gamma_S^\dagger\LF \frac{\bar{R}_{\rm dS}}{4\Mc_s^2} \RF e^{ \gamma_S\LF \frac{\bar{R}_{\rm dS}}{4\Mc_s^2} \RF  } \\
		T_{\rm NL}^1 & = \sum_{i,j,k, i\neq j\neq k} \Lc_{i} \LF \frac{2M^2}{\Mc_{s}^2} \RF  \Lc_{j} \LF \frac{2M^2}{\Mc_{s}^2} \RF  \Lc_{k} \LF \frac{2M^2}{\Mc_{s}^2} \RF \, ,\\ 
		T_{\rm NL}^2 & = \sum_{i,j,k, i\neq j\neq k} \Lc_{i} \LF \frac{2M^2}{\Mc_{s}^2} \RF  \Lc_{j} \LF \frac{2M^2}{\Mc_{s}^2} \RF \Lc_{k} \LF \frac{M^2}{\Mc_{s}^2} + \frac{\bar{ R}_{\rm dS}}{4\Mc_{s}^2}\RF \,
  \end{aligned}
  \end{equation*}
  \begin{equation}
	\begin{aligned}
		T_{\rm NL}^3 & = \sum_{i,j,k, i\neq j\neq k} \Lc_{i} \LF \frac{2M^2}{\Mc_{s}^2} \RF  \Lc_{j} \LF \frac{M^2}{\Mc_{s}^2} + \frac{\bar{ R}_{\rm dS}}{4\Mc_{s}^2} \RF \Lc_{k} \LF \frac{M^2}{\Mc_{s}^2} + \frac{\bar{ R}_{\rm dS}}{4\Mc_{s}^2}\RF\, , \\
		T_{\rm NL}^4 & = \sum_{i,j,k, i\neq j\neq k} \Lc_{i} \LF  \frac{M^2}{\Mc_{s}^2} + \frac{\bar{ R}_{\rm dS}}{4\Mc_{s}^2} \RF  \Lc_{j}\LF \frac{M^2}{\Mc_{s}^2} + \frac{\bar{ R}_{\rm dS}}{4\Mc_{s}^2} \RF \Lc_{k} \LF  \frac{M^2}{\Mc_{s}^2}+ \frac{\bar{ R}_{\rm dS}}{4\Mc_{s}^2} \RF\,. 
	\end{aligned}
	\label{tnls}
\end{equation}
Computing the amplitude of 3-point correlation, we obtain 
\begin{equation}
	A_\Rc = \sum_{i=1}^{7} B_iS_i \,, 
	\label{ampl}
\end{equation}
where 
\begin{equation}
	\begin{aligned}
		S_1 & =  2\boldsymbol{k}_1\cdot \boldsymbol{k}_2 \LT K-\frac{k_1k_2+k_2k_3+k_3k_1}{K} -\frac{k_1k_2k_3}{K^2} \RT +\textrm{perms}\, ,\\
		S_2 &  = \frac{2k_1^2k_2^2}{K} +\frac{2k_1^2k_2^2k_3}{K^2}+\textrm{perms}\,, \\
		S_3 &  \approx \,   k_3^2\LT -2K-\frac{2k_1k_2}{K}  \RT+\Cc\LF z \RF k_3^3 +\text{perms}\,\\
	\end{aligned}
\end{equation}
\begin{equation}
	\begin{aligned}
		S_4 &  = \frac{4k_1^2k_2^2k_3^2}{K^3}+\text{perms} \\
		S_5 &  =  - \frac{K^3}{3} +2Kk_1k_2+\frac{k_1k_2k_3}{3}  +\text{perms}\\
		S_6 & =  2\LF \boldsymbol{k}_1\cdot \boldsymbol{k}_2 \RF k_3^2\LT \frac{2}{K}+\frac{2k_1+2k_2}{K^2}+\frac{4k_1k_2}{K^3} \RT +\text{perms}\\
		S_7& =   \LF \boldsymbol{k}_2\cdot \boldsymbol{k}_3 \RF k_1^2k_3^2\LT -\frac{2}{K^3}-\frac{6k_2}{K^4} \RT 
	\end{aligned}
\end{equation}
where $z = \frac{K}{K_\ast}$ with $K_\ast =a_\ast H_\ast = 0.05\, {\rm Mpc}^{-1}$ is a particular reference scale and $C(z) \approx \gamma_E+\ln z -\frac{z^2}{4}+\frac{z^4}{96}$. 
Derivation of \eqref{3rda}, involves using on shell relations for the perturbed mode i.e., curvature perturbation and the background eigenvalue equation \eqref{Staroan} \cite{Koshelev:2020foq,Koshelev:2022bvg} which imply the following key equations.  
\begin{equation}
	\begin{aligned}
		\bar\square_s\Rc\approx &\, \frac{M^2}{\Mc_{s}^2}\Rc \implies & \Oc\LF\bar\square_s\RF \Rc \approx &\,\Oc\LF \frac{M^2}{\Mc_{s}^2} \RF\Rc \,\\
		\bar\square_s\Rc^{\prime}\approx &\, \LF\frac{\bar\square_s}{\Mc_s^2}+\frac{\bar{R}_{\rm dS}}{4\Mc_s^2}\RF \Rc^\prime \implies & \Oc\LF\bar\square_s\RF \Rc^\prime \approx &\, \Oc\LF \frac{M^2}{\Mc_{s}^2}+\frac{\bar{R}_{\rm dS}}{4\Mc_{s}^2} \RF\Rc^\prime\,, 
	\end{aligned}
	\label{Commtr}
\end{equation}
where $\Oc$ is an arbitrary analytic operator. In the above we used the following commutation relation in the quasi-dS approximation
 \cite{SravanKumar:2019eqt} 
\begin{equation}
	\begin{aligned}
	\nabla_{\mu}\square_s\phi  & =\square_s\nabla_{\mu}\phi-\frac{R_{\mu\nu}}{\Mc_{s}^2}\nabla^{\nu}\phi\, \\
	\implies \nabla_{\mu}F(\square_s)\phi\approx &\, \Fc\LF\square_s-\frac{R}{4\Mc_{s}^2}\RF\nabla_{\mu}\phi\,.
	\end{aligned}
\end{equation}
where $\phi$ is a scalar. 
Using \eqref{Commtr} we can bring the 3rd order action of the nonlocal gravity \eqref{NAID} into the local form, and all the effect of non-localities is transferred into the on-shell vertex factors \eqref{intr} determined by $\Tc_{\rm NL}$ and $T_{\rm NL}^1,\cdots,T_{\rm NL}^4$. From \eqref{3rda} and \eqref{intr} we can make a crucial observation that if $\bar{R}_{\rm dS} \gtrsim \Mc_s^2$ we can expect the exponential of entire function terms in \eqref{tnls} can dominate over the slow-roll suppression and eventually we get large PNGs i.e., $f_{NL}\sim \Oc(1)$.  We can see this more clearly by computing the popular limits of $f_{NL}$ called squeezed $f_{NL}^{\rm sq}$ $(k_1\to0, k_2= k_3=\frac{k}{2})$, equilateral $f_{NL}^{\rm equiv}$ $(k_1=k_2=k_3=k)$ and orthogonal $f_{NL}^{\rm orth}$ $(k_1=k_2=k/4, k_3=k/2)$ which are the useful quantities that are standard targets for CMB observations \cite{Akrami:2019izv}.
Computing $f_{\rm NL}$ for these three limits, we obtain 
\begin{equation*}
	\begin{aligned}
		f_{\rm NL }^{\rm sq}  \approx & \, \frac{5}{12} \LF 1-n_s\RF -35.5\,\Tc_{\rm NL}+8.9\,\Cc(z)\,\Tc_{\rm NL} -1.1\,\epsilon^3\,\frac{\bar{R}^2_{\rm dS}}{\Mc_s^4}\,\gamma_S^\dagger\LF \frac{\bar{R}_{\rm dS}}{4\Mc_s^2} \RF \\  &-\lambda_c\,\frac{\bar{R}_{\rm dS}}{\Mc_s^2}\, \Bigg( 5.8 \,\epsilon^2\,T_{\rm NL}^2 -  1.5\, \Cc\LF z \RF   \epsilon^2\, T_{\rm NL}^2 - 0.19 \,\epsilon^3\,  T_{\rm NL}^3\Bigg) \,,\\
		f_{\rm NL }^{\rm equiv}  \approx & \, \frac{5}{12} \LF 1-n_s\RF -46.6\,\Tc_{\rm NL}+8.9\,\Cc(z)\,\Tc_{\rm NL} -1.8\,\epsilon^3 \,\frac{\bar{R}^2_{\rm dS}}{\Mc_s^4}\,\gamma_S^\dagger\LF \frac{\bar{R}_{\rm dS}}{4\Mc_s^2} \RF \\& -\lambda_c\frac{\bar{R}_{\rm dS}}{4\Mc_s^2}\, \Bigg(7.7 \,\epsilon^2 \, T_{\rm NL}^2 - 1.5\, \Cc\LF z \RF  \, \epsilon^2\, T_{\rm NL}^2- 0.3\,\epsilon^3 \, T_{\rm NL}^3-0.02\,\epsilon^4\,  T_{\rm NL}^4\Bigg) \, .
	\end{aligned}
\end{equation*}
\begin{equation}
	\begin{aligned}
		f_{\rm NL }^{\rm ortho}  \approx & \, \frac{5}{12} \LF 1-n_s\RF -39.1\,\Tc_{\rm NL}+8.9\,\Cc(z)\,\Tc_{\rm NL} -1.2\,\epsilon^3\,\frac{\bar{R}^2_{\rm dS}}{\Mc_s^4}\,\gamma_S^\dagger\LF \frac{\bar{R}_{\rm dS}}{4\Mc_s^2} \RF \\& -\lambda_c\frac{\bar{ R}_{\rm dS}}{\Mc_{s}^2}\, \Bigg( 6.4\, \epsilon^2 \, T_{\rm NL}^2- 1.5\,  \Cc\LF z \RF \,  \epsilon^2 T_{\rm NL}^2 -0.2\, \epsilon^3\,  T_{\rm NL}^3-0.01\, \epsilon^4\, T_{\rm NL}^4 \Bigg) \, .
		\end{aligned}
		\label{fnlseo}
\end{equation}
From \eqref{fnlseo} we can witness the nonlocal corrections to the $f_{\rm NL}$ in the squeezed, equilateral and orthogonal limits. 
Further digressing the expressions \eqref{fnlseo} there are three types of nonlocal contributions we can notice. 
\begin{enumerate}
	\item Contributions containing $e^{\gamma_S\LF \frac{\bar{R}_{\rm dS}}{\Mc_s^2} \RF}$ come from the term that is quadratic in scalar curvature \eqref{listact} for the formfactor $\Fc_R\LF \square_s \RF$ in \eqref{FW}.
	\item Contributions from the cubic nonlocal scalar curvature term \eqref{listact}:  these are the vertex factors containing $T_{\rm NL}^2-T_{\rm NL}^4$. With the form factors \eqref{cubicformd} and with the conditions imposed in \eqref{entic}, we can easily deduce that
	\begin{equation}
		T_{\rm NL}^1 \ll  T_{\rm NL}^2\ll T_{\rm NL}^3\ll T_{\rm NL}^4\,
		\label{ineq}
	\end{equation}
	in the limit of $M^2\ll \Mc_s^2$ and $\bar{R}_{\rm dS}\gtrsim \Mc_s^2$. Note that since 
	$T_{\rm NL}^1$ does not depend on the ratio $\frac{\bar{R}_{\rm dS}}{\Mc_s^2}$, its contributions are far less than the contributions involving $T_{\rm NL}^2-T_{\rm NL}^4$, especially for the operators $\ell_i\LF \square_s \RF$ of the form in \eqref{entchoice}.  
	\item In the vertex factor $B_2$, the term $\Cc\LF \frac{K}{K_\ast}\RF$ comes from taking carefully the infrared (IR) limit of integration using the judicious choice $\tau= - \frac{1}{K_\ast}$ \cite{Seery:2010kh,Pajer:2016ieg}.  Usually this contribution is slow-roll suppressed in the standard single field models of inflation \cite{Pajer:2016ieg,Chen:2006nt,Burrage:2011hd}, but in our case it is modulated by analytic non-local contributions\footnote{In \eqref{fnlseo} small local contributions of the order $O\LF \epsilon^2  \RF$ are neglected. } when $\bar{R}\gtrsim \Mc_s^2$. 
\end{enumerate}
From \eqref{fnlseo} the $f_{\rm NL}$ that can be potentially probed with future observations depends on the following set of quantities
\begin{equation}
	\Bigg\{e^{\gamma_S\LF \frac{\bar{R}_{\rm dS}}{4\Mc_s^2}\RF},\,\gamma^{\dagger}_S\LF \frac{\bar{R}_{\rm dS}}{4\Mc_s^2}\RF,\,\lambda_c,\, e^{\ell_i\LF \frac{\bar{R}_{\rm dS}}{4\Mc_s^2} \RF} \Bigg\}\, .
	\label{parameterNG}
\end{equation}
As we discussed in the previous section, the scale invariance of bispectrum \eqref{ampl} is broken in the nonlocal $R^2$-like inflation due to the scale dependence of quantities \eqref{parameterNG} through 
\begin{equation}
	\Bigg\{\bar{R}_{\rm dS}\LF k \RF,\quad \Cc\LF k \RF\Bigg\}
	\label{sdrc}
\end{equation}
Due to the presence of exponentials, we can expect that the scale dependence can be significant. To quantify it we use running of $f_{\rm NL}$ defined by \cite{Chen:2006nt} 
\begin{equation}
	n_{\rm NG} \equiv \frac{d\ln f_{\rm NL}}{d\ln k}
	\label{nfnl}
\end{equation}
which we can evaluate in the various limits such as squeezed, equilateral and orthogonal. The $n_{\rm NG}$ can be evaluated using the following quantities 
\begin{equation}
	\begin{aligned}
		\frac{d f_{\rm NL }^{\rm sq} }{d\ln k} \approx & \, -\frac{dn_s}{d\ln k} +\Bigg(  -35.5+8.9\,\Cc(z) \Bigg) \frac{\pd \Tc_{\rm NL}}{\pd N} + \Cc^\dagger \LF z \RF\Bigg( 8.9\,  \Tc_{\rm NL} +\epsilon^2\lambda_c\frac{\bar{R}_{\rm dS}}{\Mc_s^2} T_{\rm NL}^2 \Bigg)\\ & -2.2\,\epsilon^4\,\frac{\bar{R}^3_{\rm dS}}{\Mc_s^6}\Bigg[\gamma_S^\dagger\LF \frac{\bar{R}_{\rm dS}}{4\Mc_s^2} \RF +\frac{1}{4} \gamma_S^{\dagger 2}\LF \frac{\bar{R}_{\rm dS}}{4\Mc_s^2} \RF +\frac{1}{4} \gamma_S^{\dagger \dagger}\LF \frac{\bar{R}_{\rm dS}}{4\Mc_s^2} \RF\Bigg] \\  &-\epsilon\,\lambda_c \frac{2\bar{R}_{\rm dS}}{\Mc_s^2} \Bigg( 5.8 \,\epsilon^2\, T_{\rm NL}^2 -  1.5\,  \epsilon^2\,\Cc\LF z \RF   T_{\rm NL}^2 - 0.76\,\epsilon^3  T_{\rm NL}^3\Bigg) \\ &
		+\epsilon\,\lambda_c \frac{\bar{R}^2_{\rm dS}}{2\Mc_s^4} \Bigg( 5.8 \,\epsilon^2\,\frac{\pd T_{\rm NL}^2}{\pd N} -  1.5\, \epsilon^2\,\Cc\LF z \RF \,   \frac{\pd T_{\rm NL}^2}{\pd N}- 0.19\,\epsilon^3 \, \frac{\pd T_{\rm NL}^3}{\pd N}\Bigg) \, ,\\
	\end{aligned}
\end{equation}
\begin{equation}
	\begin{aligned}
		\frac{df_{\rm NL }^{\rm equiv}}{d\ln k}  \approx & \, -\frac{dn_s}{d\ln k} +\Bigg(  -46.6+8.9\,\Cc(z) \Bigg) \frac{\pd \Tc_{\rm NL}}{\pd N}+  \Cc^\dagger \LF z \RF\Bigg( 8.9\,  \Tc_{\rm NL} +\epsilon^2\,\lambda_c\,\frac{\bar{R}_{\rm dS}}{\Mc_s^2}\, T_{\rm NL}^2 \Bigg) \\&
		-3.6\,\epsilon^4\,\frac{\bar{R}^3_{\rm dS}}{\Mc_s^6}\Bigg[\gamma_S^\dagger\LF \frac{\bar{R}_{\rm dS}}{4\Mc_s^2} \RF +\frac{1}{4} \gamma_S^{\dagger 2}\LF \frac{\bar{R}_{\rm dS}}{4\Mc_s^2} \RF +\frac{1}{4} \gamma_S^{\dagger \dagger}\LF \frac{\bar{R}_{\rm dS}}{4\Mc_s^2} \RF\Bigg]\\&
		-\epsilon\,\lambda_c\frac{2\bar{R}_{\rm dS}}{\Mc_s^2} \Bigg(7.7 \,\epsilon^2 \, T_{\rm NL}^2 - 1.5\,\epsilon^2\, \Cc\LF z \RF    T_{\rm NL}^2- 1.2\, \epsilon^3\,  T_{\rm NL}^3-0.12\,\epsilon^4\, T_{\rm NL}^4\Bigg) \\ &
		+\epsilon\,\lambda_c\frac{\bar{R}^2_{\rm dS}}{2\Mc_s^4} \Bigg(7.7 \epsilon^2 \frac{\pd T_{\rm NL}^2}{\pd T_{\rm NL}^2} - 1.5\,\epsilon^2\, \Cc\LF z \RF   \, \frac{\pd T_{\rm NL}^2}{\pd N}- 0.3\,\epsilon^3 \, \frac{\pd T_{\rm NL}^3}{\pd N}-0.02\,\epsilon^4  \,\frac{\pd T_{\rm NL}^4}{\pd N}\Bigg) \, , \\
		%running orth
		\frac{df_{\rm NL }^{\rm orth}}{d\ln k}  \approx & \, -\frac{dn_s}{d\ln k} +\Bigg(  -39.1+8.9\,\Cc(z) \Bigg) \frac{\pd \Tc_{\rm NL}}{\pd N}+  \Cc^\dagger \LF z \RF\Bigg( 8.9\,  \Tc_{\rm NL} +\epsilon^2\lambda_c\frac{\bar{R}_{\rm dS}}{\Mc_s^2} T_{\rm NL}^2 \Bigg) \\&
		-2.4\,\epsilon^4\,\frac{\bar{R}^3_{\rm dS}}{\Mc_s^6}\Bigg[\gamma_S^\dagger\LF \frac{\bar{R}_{\rm dS}}{4\Mc_s^2} \RF +\frac{1}{4} \gamma_S^{\dagger 2}\LF \frac{\bar{R}_{\rm dS}}{4\Mc_s^2} \RF +\frac{1}{4} \gamma_S^{\dagger \dagger}\LF \frac{\bar{R}_{\rm dS}}{4\Mc_s^2} \RF\Bigg]\\&
		-\epsilon\,\lambda_c\frac{2\bar{R}_{\rm dS}}{\Mc_s^2} \Bigg(6.4 \,\epsilon^2\, T_{\rm NL}^2 - 1.5\,\epsilon^2 \Cc\LF z \RF  \,  T_{\rm NL}^2- 0.8\,\epsilon^3\,  T_{\rm NL}^3-0.06\,\epsilon^4  \, T_{\rm NL}^4\Bigg) \\ &
		+\epsilon\,\lambda_c\frac{\bar{R}^2_{\rm dS}}{2\Mc_s^4} \Bigg(6.4\, \epsilon^2 \, \frac{\pd T_{\rm NL}^2}{\pd N} - 1.5\, \Cc\LF z \RF  \, \epsilon^2 \, \frac{\pd T_{\rm NL}^2}{\pd N}- 0.2\,\epsilon^3\,  \frac{\pd T_{\rm NL}^3}{\pd N}-0.01\,\epsilon^4 \, \frac{\pd T_{\rm NL}^4}{\pd N}\Bigg)\, .
	\end{aligned}
	\label{Rfnlseo}
\end{equation}
The relations \eqref{Rfnlseo} gives us the running of PNGs that depend on the higher derivatives of formfactors \eqref{parameterNG}. If we can probe the running of PNGs in future CMB observations, we can recontruct the structure of form factors and ultimately probe the quantum gravity. 

In the context of generalized nonlocal $R^2$-like inflation, the interesting feature is the correction to the squeezed limit of $f_{\rm NL}$ despite having single degree of freedom i.e., scalaron and slow-roll regime. This non-trivial effect is called the violation of the Maldacena consistency relation $f_{\rm NL}^{\rm eq} = \frac{5}{12}\LF 1-n_s \RF$ which we shall discuss in more detail in the next section. 
  This is truly a nonlocal effect which has no analogy with local theories. At the same time nonlocality also effects the  equilateral and orthogonal limits of $f_{NL}$ despite having the sound speed of curvature perturbation is Unity and with the adiabatic (most often called Bunch-Davies) vacuum \cite{Chen:2010xka}. In a way, it is worth to point here is that the violation of Maldacena consistency relation in GNLQG is by product of nonlocality and the curved spacetime. An important point to note here is that the $f_{\rm NL}$ \eqref{fnlseo} and \eqref{Rfnlseo} are derivations computed by summing over all the infinite terms with leading order slow-roll approximations at every step. This implies the results indicate the full nonlocal effect rather than any higher derivative. A natural question of course can be that if the large values of $f_{\rm NL}$ can be achieved in a finite higher derivative theories. First of all as it is well-known finite higher derivative theories leads to Ostrogradski instabilities due to ghost modes and are not the suitable formulations for quantum gravity with renormalizable properties around Minkowski. Suppose we assume either ghosts are heavy or assume special initial conditions for the ghosts by treating them as tachyons or Lee-Wick modes \cite{Anselmi:2018kgz}, still we cannot generate large $f_{\rm NL}$ in the finite higher derivative gravity theories unless we heavily fine tune the model parameters with unnaturally large numbers.  The GNLQG on the other hand is ghost free around Minkowski and can be super-renormalizable subjected to the arrangement of form factors with nice properties in the ultraviolet regime. 
  
\begin{figure}
	\centering
	\includegraphics[width=0.6\linewidth]{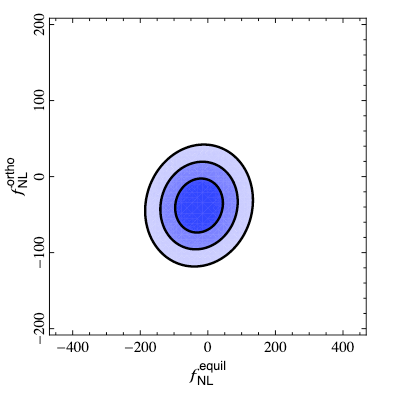}
	\caption{In the Figure we see 68\,\%, 95\,\%, and 99.7\,\% confidence regions $\LF f_{\rm NL}^{\rm eq},\,f_{\rm NL}^{\rm orth} \RF\sim O(10)$ taken from Planck 2018. The predictions of GNLQG $R^2$-like inflation lie well within the bounds in this plot along with the squeezed limit $-6<f_{\rm NL}^{\rm sq}< 4.2$.}
	\label{fig:pngpl}
\end{figure}

  Therefore, we conclude finally this section with Fig.~\eqref{fig:pngpl} which depicts the allowed range of $f_{\rm NL}$ equilateral and orthogonal limits which $R^2$-like inflation can successfully predict. Fig.~\eqref{fig:pngpl} does not represents the latest bounds on $f_{\rm NL}^{\rm sq}$ but as we discussed GNLQG does lead to detectable level of PNGs in the squeezed limit of the reduced bispectrum. 
  
\subsection{On violation of Maldacena consistency relation in GNLQG}
\label{app:violmc}

As we discussed in the previous section,  one of the prominent feature of PNGs in GNLQG is that 
 the Maldacena consistency relation can be violated and we can get $f_{\rm NL}^{\rm sq} \sim 1 $ depending on the nonlocality scale $\Mc_s$ and the choice of entire function $\gamma_S\LF \square_s \RF$ in \eqref{FW}.  In this section, we discuss how the presence of nonlocality can evade the well-known intuitively derived single field consistency theorem \cite{Maldacena:2002vr,Creminelli:2004yq,Creminelli:2011rh} which says
\begin{equation}
	f_{\rm NL}^{\rm sq} = \frac{5}{12}\LF 1-n_s \RF
\end{equation}
cannot be violated as far as we have single field, slow-roll and adiabatic initial conditions. 
The argument goes with approximating the curvature perturbation as constant on super-Hubble scales. 	The perturbed metric (say with wave-number $k_1$) in the so-called unitary gauge in which the inflaton perturbations can be put to zero \cite{Starobinsky:1982ee,Creminelli:2004yq,Creminelli:2011rh} can well estimated to be
\begin{equation}
	ds^2 = -dt^2+e^{2\zeta_{k_1}}a^2(t) dx^2\,, 
	\label{effb}
\end{equation} 
by neglecting the lapse $\Nc$ and shift functions $\Nc_i$ in the well-known Arnowitt-Deser-Misner (ADM) formalism as they are time and spatial derivatives of $\zeta_{k_1}$ \cite{Maldacena:2002vr}. Therefore, in the squeezed limit $k_1\ll k_2=k_3 = k$  \cite{Creminelli:2004yq,Creminelli:2011rh} the 3-point correlation can be Taylor expanded as 
\begin{equation}
	\begin{aligned}
		\lim_{k_1\to 0}\langle  \zeta_{\textbf{k}_1}\zeta_{\textbf{k}_2}\zeta_{\textbf{k}_3}\rangle  & \approx 
		\lim_{k_1\to 0}\langle \zeta_{\textbf{k}_1}\langle\zeta_{\textbf{k}_2}\zeta_{\textbf{k}_3}\rangle \rangle \\ &=
		-\LF 2\pi \RF^3 \delta^{3}\LF \sum_i \textbf{k}_i \RF \LF n_s-1 \RF P_{k_1}P_{k_3}
	\end{aligned} 
	\label{int-der}
\end{equation}
where $P_{k_1},\,P_{k_3}$ are the power spectrum of $k_1$ mode and $k_3$ mode respectively. The above result relies on the trick of rescaling the spatial coordinates $x^i\to e^{\zeta_{k_1}}x^i$ since $\zeta_{k_1}$ can be treated as a constant on super-Hubble scales \cite{Starobinsky:1982ee,Starobinsky:2001xq,Creminelli:2011rh}.  The reason why \eqref{int-der} cannot hold in nonlocal theory lies in the details approximations performed in this result.  

For better illustration let us consider an action of the following form which is a part of the full action \eqref{NAID}. 
\begin{equation}
	S_{R+R^2+R^3}^{\rm Non-local}  = S_{R+R^2}^{\rm local} +S_{R^3}^{\rm Non-local}
	\label{exact} 
\end{equation}
where $S_{R+R^2}^{\rm local},\, S_{R^3}^{\rm Non-local} $ are defined in \eqref{listact} with condition \eqref{LCDC}.  

The second order perturbed action of \eqref{exact} around FLRW background of \eqref{Staroan} can be computed as
\begin{equation}
	\delta^{(2)}S_{R+R^2+R^3}^{\rm Non-local} = 	\delta^{(2)}S_{R+R^2}^{\rm local}\,,
\end{equation}
where $\delta^{(2)}S_{R^3}^{\rm Non-local} \Big\vert_{\bar{	\square}\bar{R}=M^2\bar{R}} =0 $ which is the result derived in \cite{Koshelev:2022olc}. This implies the theory is completely local at the second order perturbation level of the action. This means the approximation of curvature perturbation constant on super-horizon scales perfectly holds here. 
 But obviously in this example, we get non-trivial nonlocal contributions at the 3rd order perturbed level of the action which is exactly why we get violation of consistency relation. 
%in the limit $\Fc_R\LF \square_s \RF\to f_0$. 
Clearly \eqref{exact} is a counterexample to the derivation of \eqref{int-der}. This happens because in the nonlocal case the relations \eqref{Commtr} lead to the enhancement of the interaction strength between the long wavelength mode and short wavelength modes especially when $\bar{R}\gtrsim \Mc_s^2$. 

So in summary, the proof \eqref{int-der} ideally valid only in local theories, in particular two derivative theories. In the case of nonlocal theory, one cannot locally rescale the metric and Taylor expand 3-point function because of nonlocality followed by the non-commutativity of d'Alembertian and covariant derivatives in curved spacetime \eqref{Commtr}.  

\section{Lessons from generalized nonlocal $R^2$-like inflation and its impact on the EFTs of inflationary cosmology}

The $R^2$-like inflation we get here in GNLQG totally relies on the principles of building quantum gravity that is totally different from the construction of EFTs in inflationary cosmology (EFTI) where strong assumptions are usually made \cite{Cheung:2007st,Weinberg:2008hq,Senatore:2010wk}. Here we discuss briefly the differences one can get in the context of EFTI and GNLQG nonlocal inflation (but see \cite{Koshelev:2022olc,Koshelev:2022bvg} for more detailed analysis). First of all, EFT of single inflation (EFT-SI) \cite{Cheung:2007st} prescribes that any new physics in the context of inflation has to emerge from the non-trivial sounds speeds of perturbed degrees of freedom. Surely, the context of GNLQG is a counterexample to this because the sounds speeds of perturbed modes are Unity but still we get new effects in the inflationary correlations thanks to nonlocality. This is an important point to rethink and expand the meaning of so called EFTs in cosmology.  As a simple example, let us consider the following term in the action that Weinberg has proposed  \cite{Weinberg:2008hq} in the context of EFT-SI
\begin{equation}
	\int d^4x\sqrt{-g}\Biggl[ f_1\LF \frac{\phi }{\Lambda}\RF W^{\mu\nu\rho\sigma} W_{\mu\nu\rho\sigma} \Biggr]
	\label{Weyl-Weinberg}
\end{equation} 
where $\phi$ is a canonically normalized inflaton field or the Goldstone boson in the language of EFT-SI proposed in  \cite{Cheung:2007st}. The effective scale is $\Lambda < M_p$,  and if $\phi >M_p$ that is very much natural to happen for majority of single field models of inflation (the large field ones), for example in the local $R^2$ and Higgs inflation \cite{Bezrukov:2011gp}, then the term \eqref{Weyl-Weinberg} may not be the lowest order term. Therefore, a simple generalization of \eqref{Weyl-Weinberg} would be the following
\begin{equation}
	\begin{aligned}
		\int d^4x\sqrt{-g}\Biggl[& f_1\LF \frac{\phi }{\Lambda}\RF  W^{\mu\nu\rho\sigma}\Fc_1\LF \frac{\square}{\Lambda^2},\,\frac{R}{\Lambda^2}\RF W_{\mu\nu\rho\sigma} \Biggr]
		\label{Weyl-Weinberg1}
	\end{aligned}
\end{equation} 
where $ \Fc_1\LF \frac{\square}{\Lambda^2},\,\frac{R}{\Lambda^2} \RF$ is an analytic non-polynomial functions of d'Alembertian and the Ricci scalar. We cannot truncate the terms in \eqref{Weyl-Weinberg1} especially if $R\gtrsim \Lambda^2,\,\square\gtrsim \Lambda^2$, and if $\phi\sim O\LF M_p\RF$ during inflation, one can expect $f_1\LF \frac{\phi}{\Lambda} \RF \gg 1$. We can easily notice a similarity of \eqref{Weyl-Weinberg1} with the 3rd term in the first line of GNLQG \eqref{NAID}. Suppose if we only consider the lowest order term $\frac{R}{\Mc_s^2}W_{\mu\nu\nu\sigma}W^{\mu\nu\rho\sigma}$, we can verify that we would end up with a ghost and also a non-trivial sound speed for tensor mode \cite{Baumann:2015xxa}. A catch here is that the sound speed of perturbed modes could just be an artifact of a bad truncation of the fundamental theory rather than a feature that can persist at UV completion.  Obviously there could be fundamental theories with non-trivial sound speeds all the way up to UV scales \cite{Burgess:2017ytm}.
The EFTs of multifield inflation \cite{Senatore:2010wk} are known to produce different PNGs especially in the squeezed limit. However, consistent inflationary solution in multifield inflation requires the slow-roll parameters (including the sound speeds) to be slowly varying. This would render the signature for running PNGs to be small, but in the context of PNGs in GNLQG it was found that the running of PNGs can be at least of the order of magnitude higher than those EFT models. If this is found in the future observations \cite{Meerburg:2019qqi,Karagiannis:2018jdt,Castorina:2018zfk,Munoz:2015eqa,Floss:2022grj}, we would definitely get a lead on quantum gravity research. Similarly, the predictions of running of tensor tilt in the nonlocal $R^2$-like model can be an order of magnitude higher than in the EFT frameworks of inflation \cite{Akrami:2018odb}. 
 
 In a nutshell, developments of GNLQG  does leave an impact on quantum gravity research and it highlights the importance of higher curvature terms and nonlocality in building a ghost-free quantum gravity model in curved spacetime and embedding the famous $R^2$ model of inflation into it. On the other hand, GNLQG sheds light on inflationary cosmological observables. Being precise, GNLQG not only presents a good counterexample to the the so-called EFT of inflation but also push the future observations to look for running of PNGs and running of tensor tilt in order to precisely probe early Universe physics. 

\label{sec:lessonGNLQG}

\section{Conclusions and Outlook}

It is very hard to separate cosmology and quantum gravity research as developments in each field imply understandings in the other. In this chapter, we have elaborated this in the context of $R^2$  inflation and its connection towards achieving a consistent quantum gravity. If one keeps the spirit of $R^2$ inflation and the success of Stelle gravity as a renormalizable theory, then as we have demonstrated in this chapter, a rather unavoidable route towards UV completion is presented by nonlocal higher curvature gravitational theories. We have reviewed earlier developments of the nonlocal quadratic curvature gravity (NLQG) theories which are very much possible candidates to build a consistent quantum gravity theory. We have clearly pointed out that NLQG formulation is only consistent in the flat Minkowski spacetime, so in the context of (early Universe) cosmology NLQG must be extended with higher curvature nonlocal terms. This deeply elucidates why we must in parallel aim to go forward in quantum gravity research with cosmological application. We have reviewed the recent construction of the most general theory of nonlocal gravity (up to cubic terms in curvature) that admits $R^2$-like inflation.  The generalized nonlocal quantum gravity (GNLQG) theory can be formulated with a finite parameter space by demanding ghost-freeness in the inflationary epoch. Furthermore, with appropriate assumptions on the form factors, the presence of higher curvature nonlocal terms can preserve the super-renormalizable property that was understood NLQG. This renders GNLQG as indeed a very much promising way forward where more details can be found in \cite{Koshelev:2022olc}. 

We have further elaborated in this chapter that $R^2$-like inflation in GNLQG is predictive and thus can be probed by future CMB, PGW and Large Scale Structure observations  \cite{CMB-S4:2020lpa,LiteBIRD:2022cnt,Ricciardone:2016ddg,Meerburg:2019qqi,Karagiannis:2018jdt,Castorina:2018zfk,Munoz:2015eqa,Floss:2022grj,Book:2011dz}. The takeaway message here is that GNLQG gives a new view on expected outcomes in the framework of inflation and significantly impacts our predictions of possible correlations in the present observational data as compared to the so far well explored EFTs of inflationary cosmology. Surely, there is a long way to go to understand quantum gravity. However, the content of this chapter provides new ways for investigation of GNLQG in curved spacetimes other than inflationary ones, such as black holes and anisotropic cosmology. It is also important to further understand more cosmological correlations in GNLQG involving gravitons which are under current investigation. Furthermore, GNLQG opens new doors for quantum gravity research beyond NLQG.

\label{sec:concout}

\section*{Acknowledgements}

KSK acknowledges the support from JSPS Grant-in-Aid for Scientific Research No. JP20F20320 (KAKENHI) and No. JP21H00069. KSK acknowledges the support of Royal Society in the name of Newton International Fellowship 2022. AAS was partially supported by the RSF grant 21-12-00130.

\bibliographystyle{utphys}
\bibliography{hqg.bib}

\end{document}